\documentclass[letterpaper,journal]{IEEEtran}

\usepackage{epsfig}
\usepackage{amsthm}
\usepackage[cmex10]{amsmath}
\usepackage{graphicx}
\usepackage{cite}
\usepackage{xcolor}
\usepackage{amsfonts}
\usepackage{amssymb}
\usepackage{bbm}
\usepackage{dsfont}
\usepackage{times}
\usepackage{latexsym}
\usepackage{graphicx}
\usepackage{amssymb}
\usepackage{amsfonts}
\usepackage{psfrag}
\usepackage{subfigure}
\usepackage{multicol}
\usepackage{cite}
\usepackage{setspace}
\usepackage{bigints} 

\def\bSigma{{\boldsymbol{\Sigma}}}

\def\eps{\varepsilon}

\def\ba{{\boldsymbol{a}}}

\def\bg{{\boldsymbol{g}}}

\def\bn{{\boldsymbol{n}}}

\def\bs{{\boldsymbol{s}}}

\def\bw{{\boldsymbol{w}}}
\def\bx{{\boldsymbol{x}}}
\def\by{{\boldsymbol{y}}}
\def\bz{{\boldsymbol{z}}}

\def\bA{{\boldsymbol{A}}}

\def\bF{{\boldsymbol{F}}}
\def\bG{{\boldsymbol{G}}}

\def\bI{{\boldsymbol{I}}}

\def\bP{{\boldsymbol{P}}}
\def\bQ{{\boldsymbol{Q}}}
\def\bR{{\boldsymbol{R}}}
\def\bS{{\boldsymbol{S}}}
\def\bT{{\boldsymbol{T}}}
\def\bU{{\boldsymbol{U}}}
\def\bV{{\boldsymbol{V}}}

\def\bzero{{\boldsymbol{0}}}

\def\calC{{\mathcal{C}}}

\def\calF{{\mathcal{F}}}

\def\calH{{\mathcal{H}}}

\def\calK{{\mathcal{K}}}

\def\calN{{\mathcal{N}}}

\def\calO{{\mathcal{O}}}

\def\calQ{{\mathcal{Q}}}

\def\calS{{\mathcal{S}}}

\newcommand{\comment}[1]{{{}}}

\newcommand{\showsteps}[1]{{{}}}

\newcommand{\ind}{\mathds{1}}

\newcommand{\hs}{h_{\s}}
\newcommand{\hi}{h_{\i}}

\newcommand{\mmsesu}[1]{\mathbf{mmse}_{\mathrm{su}}\brc{ #1 }}
\newcommand{\Isu}[1]{I_{\mathrm{su}}\brc{ #1}}

\newcommand{\psu}{p_{\mathrm{su}}}

\DeclareMathAlphabet{\CMmathcal}{OMS}{cmsy}{m}{n}
\renewcommand{\mathcal}[1]{\CMmathcal{#1}}
\DeclareMathAlphabet{\mathbfcal}{OMS}{cmsy}{b}{n}

\newcommand{\refeqn}[1]{(\ref{#1})}

\newcommand{\brc}[1]{\left( #1 \right)}
\newcommand{\sqbrc}[1]{\left[ #1 \right]}
\newcommand{\figbrc}[1]{\left\{ #1 \right\} }

\newcommand{\openone}{\leavevmode\hbox{\small1\normalsize\kern-.33em1}}

\newtheorem{proposition}{Proposition}

\newtheorem{corollary}{Corollary}
\theoremstyle{definition} \newtheorem{example}{Example}

\newcommand{\norm}[1]{\left\lVert{#1}\right\rVert}

\newcommand{\mmse}[1]{\langle #1\rangle}

\newcommand{\diag}{\textnormal{\textrm{diag}}}
\newcommand{\Diag}{\textnormal{\textrm{Diag}}}
\renewcommand{\j}{\mathrm{j}}
\newcommand{\E}{\textnormal{\textsf{E}}}
\newcommand{\T}{\textnormal{\textsf{T}}}
\renewcommand{\H}{\textnormal{\textsf{H}}}
\newcommand{\tr}{\mathrm{tr}}
\newcommand{\st}{\mathrm{s.t.}}
\newcommand{\til}[1]{\tilde{#1}}

\renewcommand{\d}{\; \mathrm{d}}
\newcommand{\e}{\mathrm{e}}

\newcommand{\re}{\mathrm{Re}}
\newcommand{\im}{\mathrm{Im}}

\newcommand\Real[1]{{\mathds{R}}^{#1}}
\newcommand\Complex[1]{{\mathds{C}}^{#1}}

\newcommand\bzeros[1]{{\boldsymbol{0}}_{#1}}
\newcommand\bones[1]{{\boldsymbol{1}}_{#1}}
\newcommand\bonesT[1]{{\boldsymbol{1}}_{#1}^{\T}}

\newcommand{\s}{\textnormal{\textrm{s}}}
\renewcommand{\i}{\textnormal{\textrm{i}}}

\newcommand{\Ms}{M_{\s}}
\newcommand{\Msk}[1]{M_{\s,#1}}
\newcommand{\Mi}{M_{\i}}
\newcommand{\Mil}[1]{M_{\i,#1}}

\newcommand{\rhos}{\rho_{\s}}
\newcommand{\rhosk}[1]{\rho_{\s,#1}}
\newcommand{\rhoi}{\rho_{\i}}
\newcommand{\rhoil}[1]{\rho_{\i,#1}}

\newcommand{\bHs}{\boldsymbol{H}_{\s}}
\newcommand{\bHsk}[1]{\boldsymbol{H}_{\s,#1}}

\newcommand{\bHskH}[1]{\boldsymbol{H}_{\s,#1}^{\H}}
\newcommand{\bHi}{\boldsymbol{H}_{\i}}
\newcommand{\bHil}[1]{\boldsymbol{H}_{\i,#1}}

\newcommand{\bTsk}[1]{\boldsymbol{T}_{\s,#1}}

\newcommand{\bTil}[1]{\boldsymbol{T}_{\i,#1}}

\newcommand{\bRsk}[1]{\boldsymbol{R}_{\s,#1}}

\newcommand{\bRil}[1]{\boldsymbol{R}_{\i,#1}}

\newcommand{\bWsk}[1]{\boldsymbol{W}_{\s,#1}}

\newcommand{\bWil}[1]{\boldsymbol{W}_{\i,#1}}

\newcommand{\sqRsk}[1]{\boldsymbol{R}^{1/2}_{\s,#1}}
\newcommand{\sqTsk}[1]{\boldsymbol{T}^{1/2}_{\s,#1}}
\newcommand{\sqRil}[1]{\boldsymbol{R}^{1/2}_{\i,#1}}
\newcommand{\sqTil}[1]{\boldsymbol{T}^{1/2}_{\i,#1}}

\newcommand{\bPs}{\boldsymbol{P}_{\s}}
\newcommand{\bPsk}[1]{\boldsymbol{P}_{\s,#1}}

\newcommand{\bPi}{\boldsymbol{P}_{\i}}
\newcommand{\bPil}[1]{\boldsymbol{P}_{\i,#1}}

\newcommand{\bAsk}[1]{\boldsymbol{A}_{\s,#1}}

\newcommand{\bAskH}[1]{\boldsymbol{A}_{\s,#1}^{\H}}

\newcommand{\bAil}[1]{\boldsymbol{A}_{\i,#1}}

\newcommand{\bxs}{\boldsymbol{x}_{\s}}
\newcommand{\bxsk}[1]{\boldsymbol{x}_{\s,#1}}

\newcommand{\bxskH}[1]{\boldsymbol{x}_{\s,#1}^{\H}}
\newcommand{\bxskT}[1]{\boldsymbol{x}_{\s,#1}^{\T}}
\newcommand{\bxi}{\boldsymbol{x}_{\i}}
\newcommand{\bxil}[1]{\boldsymbol{x}_{\i,#1}}

\newcommand{\bxilH}[1]{\boldsymbol{x}_{\i,#1}^{\H}}

\newcommand{\bxia}[1]{\boldsymbol{x}_{\i}^{(#1)}}

\newcommand{\bxiaT}[1]{\boldsymbol{x}_{\i}^{(#1)\T}}

\newcommand{\bxila}[2]{\boldsymbol{x}_{\i,#1}^{(#2)}}
\newcommand{\bxilaH}[2]{\boldsymbol{x}_{\i,#1}^{(#2)\H}}
\newcommand{\bxilaT}[2]{\boldsymbol{x}_{\i,#1}^{(#2)\T}}

\newcommand{\bXsk}[1]{\boldsymbol{X}_{\s,#1}}

\newcommand{\bXskH}[1]{\boldsymbol{X}_{\s,#1}^{\H}}
\newcommand{\bXi}{\boldsymbol{X}_{\i}}
\newcommand{\bXil}[1]{\boldsymbol{X}_{\i,#1}}

\newcommand{\bXilH}[1]{\boldsymbol{X}_{\i,#1}^{\H}}

\newcommand{\bvs}{\boldsymbol{v}_{\s}}
\newcommand{\bvsk}[1]{\boldsymbol{v}_{\s,#1}}

\newcommand{\bvsT}{\boldsymbol{v}_{\s}^{\T}}

\newcommand{\bvia}[1]{\boldsymbol{v}_{\i}^{(#1)}}

\newcommand{\bviaT}[1]{\boldsymbol{v}_{\i}^{(#1)\T}}

\newcommand{\bvila}[2]{\boldsymbol{v}_{\i,#1}^{(#2)}}

\newcommand{\bQsk}[1]{\boldsymbol{Q}_{\s,#1}}

\newcommand{\bQil}[1]{\boldsymbol{Q}_{\i,#1}}

\newcommand{\calQs}{\calQ_{\s}}
\newcommand{\calQi}{\calQ_{\i}}

\newcommand{\tilQsk}[1]{\til{\boldsymbol{Q}}_{\s,#1}}

\newcommand{\tilQil}[1]{\til{\boldsymbol{Q}}_{\i,#1}}

\newcommand{\tcalQ}{\til{\calQ}}

\newcommand{\qsk}[1]{q_{\s,#1}}

\newcommand{\tilqsk}[1]{\til{q}_{\s,#1}}
\newcommand{\tilpsk}[1]{\til{p}_{\s,#1}}
\newcommand{\qil}[1]{q_{\i,#1}}
\newcommand{\pil}[1]{p_{\i,#1}}
\newcommand{\tilqil}[1]{\til{q}_{\i,#1}}
\newcommand{\tilpil}[1]{\til{p}_{\i,#1}}

\newcommand{\betask}[1]{\beta_{\s,#1}}

\newcommand{\betail}[1]{\beta_{\i,#1}}

\newcommand{\funcG}{G^{(u)}}
\newcommand{\measure}{\mu^{(u)}}
\newcommand{\mgf}{M^{(u)}}
\newcommand{\funcRate}{I^{(u)}}
\newcommand{\funcT}[1]{T^{(u)}_{#1}}

\newcommand{\bzsk}[1]{\boldsymbol{z}_{\s,#1}}

\newcommand{\bzil}[1]{\boldsymbol{z}_{\i,#1}}

\newcommand{\bzilH}[1]{\boldsymbol{z}_{\i,#1}^{\H}}

\newcommand{\epssk}[1]{\eps_{\s,#1}}
\newcommand{\epsil}[1]{\eps_{\i,#1}}
\newcommand{\xisk}[1]{\xi_{\s,#1}}
\newcommand{\xiil}[1]{\xi_{\i,#1}}
\newcommand{\xiilbar}[1]{\overline{\xi}_{\i,#1}}
\newcommand{\epsilbar}[1]{\overline{\eps}_{\i,#1}}

\newcommand{\barzil}[1]{\overline{\boldsymbol{z}}_{\i,#1}}

\newcommand{\barAsk}[1]{\bar{\boldsymbol{A}}_{\s,#1}}

\newcommand{\barAil}[1]{\bar{\boldsymbol{A}}_{\i,#1}}

\newcommand{\barBil}[1]{\bar{\boldsymbol{B}}_{\i,#1}}

\newcommand{\bcalGs}{\mathbfcal{G}_{\s}}

\newcommand{\bLam}{\boldsymbol{\Lambda}}

\newcommand{\cf}{\textrm{cf.}\ }
\newcommand{\eg}{\textit{e.g.}}

\newcommand{\ie}{\textit{i.e.}}

\newcommand{\perse}{\textit{per se }}
\newcommand{\via}{\textit{via }}

\newcommand{\vide}{\textit{vide }}

\newcommand{\vs}{\textit{vs.}\;}

\newcommand{\bGsk}[1]{\boldsymbol{G}_{\s,#1}}

\newcommand{\bGskH}[1]{\boldsymbol{G}_{\s,#1}^{\H}}

\newcommand{\bGil}[1]{\boldsymbol{G}_{\i,#1}}

\newcommand{\bGilH}[1]{\boldsymbol{G}_{\i,#1}^{\H}}

\newcommand{\bssk}[1]{\boldsymbol{s}_{\s,#1}}

\newcommand{\bsil}[1]{\boldsymbol{s}_{\i,#1}}

\newcommand{\bEsk}[1]{\boldsymbol{E}_{\s,#1}}

\newcommand{\bFsk}[1]{\boldsymbol{F}_{\s,#1}}

\renewcommand{\phi}{\varphi}

\newcounter{MYtempeqncnt}

\renewcommand{\l}{\ell}

\begin{document}
\IEEEoverridecommandlockouts

\title{Large-System Analysis of Correlated MIMO Multiple Access Channels with Arbitrary Signaling in the Presence of Interference}

\author{
\authorblockN{Maksym~A.~Girnyk, Mikko Vehkaper\"{a}, Lars K. Rasmussen\\}
\thanks{The research leading to these results has received funding from the European Research Council under the European Community's Seventh Framework Programme (FP7/2007-2013) / ERC grant agreement no 228044. The work has further been supported in parts by the ARC Grant DP0986089, and the VR grant 621-2009-4666. The material of this paper was presented in parts at the IEEE GLOBECOM Conference 2012, Anaheim, U.S.A., Dec. 2012.}
\thanks{The authors are with the School of Electrical Engineering and the ACCESS Linnaeus Centre at KTH Royal Institute of Technology, Stockholm, Sweden (e-mail: \{mgyr, mikkov, lkra\}@kth.se).}
}

\maketitle

\begin{abstract} \label{sec:abstract}
Presence of multiple antennas on both sides of a communication channel promises significant improvements in system throughput and power efficiency. In effect, a new class of large multiple-input multiple-output (MIMO) communication systems has recently emerged and attracted both scientific and industrial attention. To analyze these systems in realistic scenarios, one has to include such aspects as co-channel interference, multiple access and spatial correlation. In this paper, we study the properties of correlated MIMO multiple-access channels in the presence of external interference. Using the replica method from statistical physics, we derive the ergodic sum-rate of the communication for arbitrary signal constellations when the numbers of antennas at both ends of the channel grow large. Based on these asymptotic expressions, we also address the problem of sum-rate maximization using statistical channel information and linear precoding.
The numerical results demonstrate that when the interfering terminals use discrete constellations, the resulting interference becomes easier to handle compared to Gaussian signals. Thus, it may be possible to accommodate more interfering transmitter-receiver pairs within the same area as compared to the case of Gaussian signals. In addition, we demonstrate numerically for the Gaussian and QPSK signaling schemes that it is possible to design precoder matrices that significantly improve the achievable rates at low-to-mid range of signal-to-noise ratios when compared to isotropic precoding.

\end{abstract}

\section{Introduction}\label{sec:intro}
\begin{figure}
\centering
  \psfrag{xs1}[][][0.9]{\textcolor{blue}{$\bxsk{1}$}}
  \psfrag{xsK}[][][0.9]{\textcolor{blue}{$\bxsk{K}$}}
  \psfrag{Gs1}[][][0.9]{\textcolor{blue}{$\bGsk{1}$}}
  \psfrag{GsK}[][][0.9]{\textcolor{blue}{$\bGsk{K}$}}
  \psfrag{Hs1}[][][0.9]{\textcolor{blue}{$\bHsk{1}$}}
  \psfrag{HsK}[][][0.9]{\textcolor{blue}{$\bHsk{K}$}}
  \psfrag{xi1}[][][0.9]{\textcolor{red}{$\bxil{1}$}}
  \psfrag{xiL}[][][0.9]{\textcolor{red}{$\bxil{L}$}}
  \psfrag{Gi1}[][][0.9]{\textcolor{red}{$\bGil{1}$}}
  \psfrag{GiL}[][][0.9]{\textcolor{red}{$\bGil{L}$}}
  \psfrag{Hi1}[][][0.9]{\textcolor{red}{$\bHil{1}$}}
  \psfrag{HiL}[][][0.9]{\textcolor{red}{$\bHil{L}$}}
  \psfrag{y}[][][0.9]{$\by$}
  \psfrag{U1}[][][0.8]{\textcolor{blue}{User 1}}
  \psfrag{UK}[][][0.8]{\textcolor{blue}{User $K$}}
  \psfrag{I1}[][][0.8]{\textcolor{red}{Int. 1}}
  \psfrag{IL}[][][0.8]{\textcolor{red}{Int. $L$}}
  \psfrag{R}[][][0.8]{Rx}
\includegraphics[width=\linewidth]{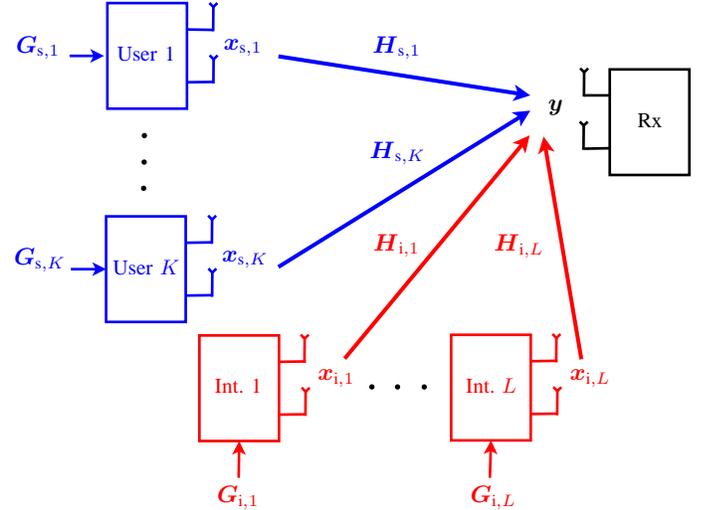}
  \caption{MIMO multiple-access channel in the presence of interference.}
  \label{fig:mac}
\end{figure}

During the last decade, multi-antenna communications has received an increased interest both from academia and industry. Pioneering research
by Foschini, Gans and Telatar \cite{foschini1998limits,telatar1999capacity} on the topic suggested that the new class of multiple-input multiple-output (MIMO) systems allowed the transmission rate to be increased roughly linearly in the number of antennas available at the transmitter and receiver.  Measurements both indoors~\cite{werner2012lte} and outdoors~\cite{oikonomopoulos2012outdoor} have also confirmed the throughput gains of the multi-antenna transmission.

The main price to pay for the benefits offered by multi-antenna transmission is the hardware and signal processing complexity at both the transmitter and receiver. It is therefore of great importance to analyze the potential performance gains of MIMO processing in realistic scenarios before employing the techniques in practice.  For example, in the uplink of a cellular system, the effects of \emph{co-channel interference} emerging from other cells need to be taken into account.  As observed in~\cite{Blum2003interference}, such interference has also a surprising influence on the optimal power allocation strategy at the transmitter.
In addition to co-channel interference,
\emph{spatial correlation}~\cite{tulino2005impact} and the type of channel inputs~\cite{muller2003channel} have a great impact on the achievable rate of the channel.  However, analysis of realistic scenarios tend to be mathematically challenging and numerical simulations are time consuming, especially if discrete signaling is employed at the transmitter.  Some simplifying assumptions are therefore needed to make the problem tractable.

Asymptotic approaches developed within the field of \emph{random matrix theory} greatly facilitate the analysis of achievable ergodic rates (mutual information averaged over channel realizations) in MIMO systems. Such methods were used already in the early works \cite{foschini1998limits, telatar1999capacity} to assess the capacity of multi-antenna transmission.  At the same time, several approaches using random matrix theory for the analysis of the spectral efficiency of large code division multiple access (CDMA) systems \cite{grant1998random, verdu1999spectral, tse1999linear,
rapajic2000information, evans2000linearmud} were reported.  The multi-antenna results were later extended to the case of spatial correlation in~\cite{chuah2002capacity} and then to \emph{MIMO multiple-access channel} (MIMO-MAC) in~\cite{couillet2011deterministic} (\vide Fig.~\ref{fig:mac}).
Some analysis of MIMO systems in the case of co-channel interference have also been carried out under the assumption of Gaussian channel inputs. The first analytical results using random matrix theory were obtained in~\cite{lozano2002capacity} assuming uncorrelated channels and interferers.  Some later efforts
\cite{kang2007capacity,wang2009interference,chiani2010interference}
have extended this analysis to different assumptions about correlations and numbers of antennas present in the system.

Although the above methods assume formally that the system size grows without bound, they usually provide a good approximation also for the performance of finite-sized systems. Furthermore, the underlying large-system assumption \perse has recently found a practical application in so-called
\emph{massive MIMO}~\cite{marzetta2010noncooperative,huh2012massive,rusek2013scaling,larsson2013massive}. Typically this concept entails a multiuser system where a single base station, equipped with a very large antenna array, is used to serve a smaller number of terminals simultaneously. Apart from the aforementioned throughput gains, such systems allow for a significant reduction of the transmit power and reduced-complexity signal processing~\cite{rusek2013scaling}. Consequently, asymptotic random-matrix methods have been widely applied for the analysis of various aspects of massive MIMO~\cite{hoydis2013massive,muller2013blind,zhang2013capacity} systems.

The aforementioned studies regarding MIMO systems with co-channel interference have all concentrated on the special case, where Gaussian signals are transmitted both by the desired user and the interfering terminals. This is in contrast to real-world systems, where discrete constellations such as QPSK and QAM are used. These realistic cases are, however, out-of-bounds for random matrix theory, except for setups where sub-optimal linear detection and per-stream decoding is considered.
To investigate the performance bounds of generic systems with \emph{non-Gaussian channel inputs}, a tool borrowed from the field of statistical physics, namely the \emph{replica method}, has been recently used.

The replica method was invented by Kac~\cite{kac1968toeplitz}, and is widely known due to its early applications to \emph{spin glasses}
\cite{edwards1975theory,sk1975spinglasses}.
The replica framework provides a powerful set of mathematical tools for computing average quantities within large many-body systems and has since been applied to various problems in science and engineering.
In the context of information theory, it was used to assess the spectral efficiency of large CDMA systems with antipodal signaling by Tanaka in~\cite{tanaka2001statistical,tanaka2002statistical}. Later, Guo and Verd\'{u} generalized the approach to CDMA with arbitrary signaling~\cite{guo2005cdma}.  Meanwhile, in~\cite{muller2003channel} and~\cite{takeda2007statistical} the method was applied to spatially correlated MIMO channels with binary inputs. These works were further generalized in~\cite{wen2007asymptotic} to the analysis of the sum-rate of a MIMO-MAC. A somewhat different approach was taken in~\cite{moustakas2003mimo}, where the replica method was used to analyze the moments of mutual information of a MIMO system with co-channel interference. The results in~\cite{moustakas2003mimo} were obtained, however, under the assumption of Gaussian signaling at all terminals.

In the present paper we extend our previous work~\cite{girnyk2012asymptotic} and investigate
the performance and sum-rate maximization of a correlated MIMO-MAC using the replica method.  The analysis encompasses
the presence of correlated non-Gaussian interferers and arbitrary inputs at all terminals.  As in
\cite{tanaka2001statistical,guo2005cdma,tanaka2002statistical,takeda2007statistical,wen2007asymptotic,moustakas2003mimo},
the results are obtained
under the technical assumption of replica symmetric ansatz. To summarize, the following contributions are reported:
\begin{itemize}
\item We derive an expression for the asymptotic sum-rate of the MIMO-MAC with spatial correlation and in the presence of spatially correlated multi-antenna interferers. The analysis is valid for arbitrary channel inputs at all terminals\footnote{Throughout the paper, we assume that the signaling scheme used at each interfering terminal is known to the receiver. In the cellular setting, such information can be exchanged between neighboring base stations \via an existent backhaul link with a very small overhead.  Alternatively, our results provide an upper bound to the setting where the signaling schemes are not known or the base station is misinformed about them.} and is carried out in the \emph{large-system limit} (LSL), where the numbers of antennas at both ends of each MIMO channel grow without bound at a constant rate. As expected, several prior results are obtained as special cases of the analysis. For instance, in the absence of interferers, our results degenerate to those reported in~\cite{aktas2006scaling,couillet2011deterministic}
when Gaussian signals are employed and to those provided in~\cite{wen2007asymptotic} when arbitrary signal constellations are used. Finally, in the presence of interference and under assumption of Gaussian signaling, our results partly reduce to the expression of the mean mutual information derived in~\cite{moustakas2003mimo}.
\item We address the precoder optimization problem for both Gaussian and finite-alphabet signaling schemes under the assumptions of full channel state information (CSI) at the receiver\footnote{In practice, the CSI at the receiver is estimated by using known training sequences sent by the users.  In the cellular setting, the same procedure can also be used to estimate the channels of the interferers if there is sufficient synchronization and the base station is informed about the set of training sequences that are in use in the neighboring cell.  The latter can be achieved with a small overhead in the backhaul link.  Furthermore, in order to implement soft handover between the cells, the base stations tend to establish tight synchronization with the users from other cells close to the cell border. Hence, estimating the interfering users' channels is possible also in practical systems.  The mathematical model considered in this paper provides thus an upper bound for the performance of a practical systems that estimates the CSI or has only partial knownledge of the latter.} and \emph{statistical CSI at the transmitter}. By using the asymptotic sum-rate as an objective function for the corresponding optimization problem, we obtain the precoding matrices for each user.
\end{itemize}

The remainder of the paper is organized as follows. In the following section, we describe the system model and formulate the main problem. In addition, we discuss the necessary details regarding the MIMO channels with perfect CSI at the transmitter. Next, in Section~\ref{sec:sumRate}, we present the main result of the paper, that is, the asymptotic sum-rate of a MIMO-MAC in the presence of interference. Section~\ref{sec:optimizationCovariance} then addresses the precoder optimization problem, followed by Section~\ref{sec:numericalResults}, where we present numerical results and discussion. Finally, in Section~\ref{sec:conclusion}, we conclude the paper. The proofs are relegated to the appendices.

\emph{Notation:}
Throughout this paper we will use upper case bold-faced letters to denote matrices, \eg, $\bA$, with elements denoted by $[\bA]_{i,j}$, lower case bold-faced letters to denote column vectors, \eg, $\ba$, with elements $a_i$, and lower case light-faced letters to denote scalar variables, \eg, $a$. Superscripts $\T$ and $\H$ denote transpose and Hermitian adjoint operators, respectively. Meanwhile, $\bA^{1/2}$, $\tr\{\bA\}$ and $\det(\bA)$ denote the principal square root, the trace and the determinant of matrix $\bA$. We differentiate between operators $\diag(\ba)$, which denotes a diagonal matrix containing the coefficients of vector $\ba$ on its main diagonal, and $\Diag(\bA)$, which denotes a column vector containing the diagonal entries of matrix $\bA$. Also, $\bI$, $\bzeros{}$ and $\bones{}$ denote the identity matrix, the zero matrix and the all-ones vector of appropriate sizes. Operator $\E\{\cdot\}$ denotes the expectation, $\ind(\cdot)$ denotes the indicator function, $\otimes$ represents the Kronecker product, and $\bA \succeq \bzeros{}$ implies that the matrix $\bA$ is positive semidefinite. Finally, $\re\{\cdot\}$ and $\im\{\cdot\}$ stand for the real and imaginary parts of the argument.

\section{Preliminaries}\label{sec:preliminaries}
\subsection{System Model}
\label{sec:systemModel}
Consider the scenario where $K$ multi-antenna terminals communicate to a single multi-antenna receiver in the presence of $L$ multi-antenna interferers, as depicted in Fig.~\ref{fig:mac}. An uplink cellular communication system in the presence of inter-cell interference can be regarded as an example of such a scenario. The numbers of antennas at transmitter $k$, interferer $\l$ and the receiver, are denoted by $\Msk{k}$, $\Mil{\l}$, and $N$, respectively. The discrete-time received vector is given by
\begin{equation} \label{eqn:signalReceived}
    \by = \sum_{k=1}^K \bHsk{k} \bxsk{k} + \sum_{\l=1}^L \bHil{\l} \bxil{\l} + \bn,
\end{equation}
where $\bxsk{k} \in \Complex{\Msk{k}}$ is the zero-mean transmitted signal vector of the $k$th user with covariance matrix $\E\{\bxsk{k} \bxskH{k}\} = \bPsk{k}$ and $\bxil{\l} \in \Complex{\Mil{\l}}$ is the transmitted signal vector of the $\l$th interferer having zero-mean and covariance $\E\{\bxil{\l} \bxilH{\l}\} = \bPil{\l}$.  To satisfy long-term power constraints at the transmitters, we require that $\tr\{\bPsk{k}\} \leq \Msk{k}$ and $\tr\{\bPil{\l}\} \leq \Mil{\l}$.  The noise vector $\bn \in \Complex{N}$ has independent circularly symmetric complex Gaussian (CSCG) entries with unit variance. Matrices $\bHsk{k} \in \Complex{N \times \Msk{k}}$ and $\bHil{\l} \in \Complex{N \times \Mil{\l}}$ denote the MIMO channels between user $k$ and the receiver and between interferer $\l$ and the receiver, respectively. The channels are assumed to be flat-fading and are modeled \via the Kronecker model~\cite{chizhik2000effect}, that is
\begin{subequations}\label{eqn:kronecker}
\begin{align}
    \bHsk{k} =& \sqrt{\frac{\rhosk{k}}{\Msk{k}}} \sqRsk{k} \bWsk{k} \sqTsk{k},\\
    \bHil{\l} =& \sqrt{\frac{\rhoil{\l}}{\Mil{\l}}} \sqRil{\l} \bWil{\l} \sqTil{\l},
\end{align}
\end{subequations}
where $\rhosk{k}$ and $\rhoil{\l}$ represent average signal-to-noise ratios (SNRs) of the corresponding links and matrices $\bWsk{k}$ and $\bWil{\l}$ have i.i.d. CSCG entries of unit variance.  The correlation matrices at the receive end are denoted by $\bRsk{k}$ and $\bRil{\l}$, while $\bTsk{k}$ and $\bTil{\l}$ represent the correlation matrices at the transmit end of the corresponding channels. To ensure that the correlation matrices do not influence the average path gains, they are normalized as
\begin{subequations}\label{eqn:correlationNormalized}
\begin{align}
    \tr\{\bRsk{k}\} =&\; N, \quad \tr\{\bTsk{k}\}= \Msk{k}, \\
    \tr\{\bRil{\l}\} =&\; N, \quad \tr\{\bTil{\l}\}= \Mil{\l}.
\end{align}
\end{subequations}

For later convenience, we write the input covariance matrices
in terms of two \emph{precoder matrices}, that is, $\bPsk{k} = \bGsk{k} \bGskH{k}$ and $\bPil{\l} = \bGil{\l} \bGilH{\l}$,  and let them depend on  the statistical CSI, \ie, the knowledge of $\{\rhosk{k},\bTsk{k},\bRsk{k}\}$ and $\{\rhoil{\l},\bTil{\l},\bRil{\l}\}$, respectively.
Thus, denoting $\bssk{k}$ and $\bsil{\l}$ for independent vectors with i.i.d. zero-mean unit variance entries, we may write $\bxsk{k} = \bGsk{k} \bssk{k}$ and $\bxil{\l} = \bGil{\l} \bsil{\l}$ without loss of generality.
  This formulation is especially useful when we consider the optimization of the input covariance for discrete signals.
For notational simplicity, we also denote $\Ms \triangleq \sum_{k=1}^K \Msk{k}$ and $\Mi \triangleq \sum_{\l=1}^L \Mil{\l}$, and rewrite the input-output relation of the resulting MIMO channel as
\begin{equation} \label{eqn:matrixChannel}
    \by = \bHs \bxs + \bHi \bxi + \bn,
\end{equation}
where $\bHs \triangleq \sqbrc{\bHsk{1}, \ldots, \bHsk{K}} \in \Complex{N \times \Ms}$, $\bHi\triangleq\sqbrc{\bHil{1}, \ldots, \bHil{L}} \in \Complex{N \times \Mi}$, $\bxs \triangleq [\bxsk{1}^{\T}, \ldots, \bxsk{K}^{\T}]^{\T} \in \Complex{\Ms}$, $\bxi \triangleq [\bxil{1}^{\T}, \ldots, \bxil{L}^{\T}]^{\T} \in \Complex{\Mi}$.

\subsection{Problem Statement}
\label{sec:problemStatement}
Define the instantaneous CSI at the receiver as $\calH~\triangleq~\{\bHs, \bHi\}$.  Given that all channels are ergodic and the receiver knows $\calH$, the distribution of $\bxs$, as well as the distribution of $\bxi$, we can write down the average mutual information
\begin{equation}\label{eqn:mutualInfo}
         I(\by; \bxs) =  h\brc{\by|\calH} - h\brc{\by|\bxs,\calH},
\end{equation}
where the differential entropy terms are given by
\begin{subequations}
\begin{align}
    h\brc{\by|\calH} =& -\E_{\by,\calH} \ln \E_{\bxs,\bxi} p\brc{\by|\bxs,\bxi,\calH}, \label{eqn:entropyY}\\
    h\brc{\by|\bxs,\calH} =& - \E_{\by,\bxs,\calH} \ln \E_{\bxi} p\brc{\by|\bxs,\bxi,\calH} \label{eqn:entropyYConditional},
\end{align}
\end{subequations}
and the conditional distribution of the channel~\refeqn{eqn:matrixChannel} reads
\begin{equation}\label{eqn:distributionChannel}
         p\brc{\by|\bxs,\bxi,\calH} = \frac{1}{\pi^N}\e^{-\norm{\by-\bHs \bxs - \bHi \bxi}^2}.
\end{equation}
The mutual information in~\refeqn{eqn:mutualInfo} represents an achievable sum-rate of the MIMO-MAC~\refeqn{eqn:signalReceived} when the receiver does not decode the interference signal $\bxi$.  Given statistical knowledge of the channels, the sum-rate could then, in principle, be maximized by designing the precoder matrices $\bGsk{k},\;\forall k$. Unfortunately, the explicit expression for~\refeqn{eqn:mutualInfo} is not known in general. Moreover, its numerical evaluation is computationally expensive due to averaging of~\refeqn{eqn:entropyY} and~\refeqn{eqn:entropyYConditional} over the channel realizations.  Even more serious difficulty arises when the data symbols are non-Gaussian. In this case, one needs to compute two sums over an exponential number (w.r.t.\ the numbers of transmit antennas and bits in the constellation) of terms for every realization of the channel.

The aim of the present paper is to find a computationally feasible expression for the ergodic mutual information~\refeqn{eqn:mutualInfo} given arbitrary channel inputs.
To make the analysis tractable, we consider the asymptotic regime where the system size grows large and use the replica method to compute the individual entropy terms.  These expressions (\vide Section~\ref{sec:sumRate}) are then used to optimize the covariance matrices so that the mutual information is maximized (\vide Section~\ref{sec:optimizationCovariance}).  Finally, the large system result is used to give an approximation for the original quantity~\refeqn{eqn:mutualInfo} when the system size is finite.

\subsection{Mutual Information and the MMSE of a Fixed MIMO Channel}
\label{sec:channelFixed}
To finish this section, we discuss the problem of finding the mutual information and the minimum mean squared error (MMSE) of a fixed MIMO channel. These results are used later in the paper to evaluate the asymptotic mutual information obtained \via the replica method.

\subsubsection{General Case}
\label{sec:channelFixedGeneral}
Consider the following multi-antenna communication channel
\begin{equation}\label{eqn:channelFixed}
    \bz = \bA \bx + \bw,
\end{equation}
where $\bA$ is a fixed $N \times M$ channel matrix and $\bw \in \Complex{N}$ has i.i.d. CSCG elements of unit variance.  The channel inputs $\bx = \bG \bs \in \Complex{M}$ are a combination of a precoder matrix $\bG$ and vector $\bs$ that has i.i.d. zero-mean unit variance entries, with constraint $\tr\figbrc{\bP} \leq M$ where $\E\{\bx\bx^{\H}\} = \bP$.  The conditional distribution of this fixed \emph{single-user} (su) channel%
\footnote{In this and the following sections,
the probabilities related to the channel~\eqref{eqn:channelFixed} are denoted $\psu$.  This is to make a clear separation to the probabilities related to the original channel~\eqref{eqn:matrixChannel}.
}
is given by
\begin{equation}\label{eqn:channelFixedDistribution}
    \psu\brc{\bz|\bx,\bA} = \frac{1}{\pi^M} \e^{-\norm{\bz-\bA\bx}^2}.
\end{equation}
The posterior mean estimate of $\bx$ is denoted $\mmse{\bx} \triangleq \E \figbrc{\bx|\bz,\bA}$, where the expectation is taken over the posterior density $\psu\brc{\bx|\bz,\bA}$ obtained from the prior distribution $p(\bx)$ and~\refeqn{eqn:channelFixedDistribution} \via Bayes' theorem.
For future convenience, the conditional \emph{MMSE matrix} is defined here through parametrization
\begin{align}\label{eqn:matrixMmse}
    \mmsesu{\bx, \bA}\triangleq
    \E_{\bz,\bx}\figbrc{ (\bx - \mmse{\bx})(\bx - \mmse{\bx})^{\H}} \in \Complex{M \times M},
\end{align}
where the expectation is w.r.t.\ the joint distribution
$\psu\brc{\bz,\bx|\bA}$.
Similarly, the mutual information reads
\begin{equation}\label{eqn:mutualInfoGeneric}
\Isu{\bz; \bx|\bA} =
\E_{\bz,\bx} \ln \psu \brc{\bz|\bx,\bA}
- \E_{\bz} \ln \E_{\bx} \,\psu \brc{\bz|\bx,\bA},
\end{equation}
where the expectations are again w.r.t.\ $\psu\brc{\bz,\bx|\bA}$.

\begin{figure*}
\normalsize
\setcounter{MYtempeqncnt}{\value{equation}}
\setcounter{equation}{26}
\begin{equation} \label{eqn:mmse16qam}
    \E_{z_m,x_m} \figbrc{|x_m-\mmse{x_m}|^2|a_m} = g_m^2 - \frac{g_m^2}{10\sqrt{\pi}} \bigintsss_{\mathbb{R}} \frac{\brc{3 \e^{-\frac{4 g_m^2 a_m^2}{5}} \sinh\brc{\frac{6 g_m a_m}{\sqrt{10}}s} + \sinh\brc{\frac{2 g_m a_m}{\sqrt{10}}s} }^2}{\e^{-\frac{4 g_m a_m}{5}} \cosh \brc{\frac{6 g_m a_m}{\sqrt{10}}s} + \cosh\brc{\frac{2g_m a_m}{\sqrt{10}}s} } \; \e^{-\frac{10s^2 + g_m^2 a_m^2}{10} } \d s,
\end{equation}
\hrulefill
\setcounter{equation}{\value{MYtempeqncnt}}
\end{figure*}

Below we present two important special cases and provide the corresponding expressions for the mutual information and the MMSE.

\begin{example}[\emph{Gaussian inputs}]
The MMSE detector becomes linear if the channel input $\bx$ is a CSCG vector given $\bG$. In this case, the output of the MMSE detector reads
\newline
\begin{align}\label{eqn:mmseEstimate}
         &\mmse{\bx} =  \brc{\bP^{-1} + \bA^{\H}\bA}^{-1} \bA^{\H} \bz,
\end{align}
and the MMSE matrix is given by
\begin{align}\label{eqn:mmse}
         &\mmsesu{\bx, \bA} =\brc{\bP^{-1} + \bA^{\H} \bA}^{-1}.
\end{align}
The mutual information reduces also to the well-known formula
\begin{equation}\label{eqn:mutualInfoGauss}
         \Isu{\bz; \bx|\bA} =  \ln\det\brc{\bI_N + \bA \bP \bA^{\H}}.
\end{equation}
\end{example}

\begin{example}[\emph{Discrete channel inputs}]
Let the entries of $\bs$ be independently drawn from a discrete constellation (\eg, QPSK, QAM) of cardinality $C$, so that
$\bs$ is uniformly distributed over the set $\{\bs_1,\ldots,\bs_{C^M}\}$.
 For fixed $\bG$ we may then treat also $\bx$ as being uniformly drawn from a set $\{\bx_1,\ldots,\bx_{C^M}\}$ of possible input vectors. Denoting
\begin{align}\label{eqn:outputDiscrete}
         \psu(\bz|\bA) =&  \frac{1}{C^M} \sum_{i=1}^{C^M} \frac{1}{\pi^M} \e^{-\norm{\bz - \bA \bx_i}^2},
\end{align}
the MMSE estimate of $\bx$ is by definition given as
\begin{equation}\label{eqn:estimateMmseDiscrete}
         \mmse{\bx} =\frac{1}{{C^M}} \sum_{i=1}^{C^M}\frac{\bx_i \; \psu(\bz| \bx_i,\bA)}{\psu(\bz|\bA)}.
\end{equation}
The MMSE matrix is thus obtained from
\begin{equation}
\mmsesu{\bx, \bA}= \bP
- \int \mmse{\bx} \mmse{\bx}^{\H} \psu(\bz|\bA)  d \bz,
\end{equation}
while the mutual information between $\bz$ and $\bx$ reads
\begin{align}\label{eqn:mutualInfoDiscrete}
         & \Isu{\bz; \bx|\bA,\psu} = N + M \ln C \nonumber\\
         &\qquad - \frac{1}{C^M} \sum_{i=1}^{C^M} \E_{\bw} \Bigg\{ \ln \sum_{j=1}^{C^M} \e^{-\norm{\bA(\bx_i - \bx_j) + \bw}^2} \Bigg\} .
\end{align}
\end{example}

\subsubsection{Parallel Gaussian Channels}
\label{sec:channelFixedDiagonal}

Assume that the channel matrix $\bA$ is diagonal with real-valued entries $a_1, \ldots, a_M$.  Let $\bG$ also be a real diagonal matrix formed by $g_1, \ldots, g_M$ so that the MIMO channel~\refeqn{eqn:channelFixed} decouples into a bank of parallel channels
\begin{equation}\label{eqn:channelFixedUncorrelated}
    z_m = a_m x_m + w_m.
\end{equation}
The MMSE estimate of $x_m$ for the $m$th channel is then $\mmse{x_m} = \E \figbrc{x_m|z_m,a_m}$, while the MMSE matrix is diagonal with
$\E_{z_m,x_m} \figbrc{|x_m-\mmse{x_m}|^2|a_m}$ as its $(m,m)$th element.
The mutual information \eqref{eqn:mutualInfoGeneric} reduces in this case to a form
\begin{equation}\label{eqn:mutualInfoParallel}
\Isu{\bz; \bx|\bA} = \sum_{m=1}^{M} I\brc{z_m;x_m|a_m,\psu}.
\end{equation}
In the following examples we discuss three analytically tractable special cases, which will be useful later in Section~\ref{sec:numericalResults}.
\begin{example}[\emph{Gaussian inputs}]
In this scenario, we have $x_m\sim\calC\calN(0,g^{2}_m)$ so that the MMSE estimate of $x_m$ becomes
\begin{equation}\label{eqn:estimateMmseGaussian}
    \mmse{x_m} = \frac{g_m^2 a_m z_m}{1 + g_m^2 a_m^2},
\end{equation}
leading to
\begin{equation}\label{eqn:matrixMmseGaussian}
    \E_{z_m,x_m} \figbrc{|x_m-\mmse{x_m}|^2|a_m}
    = \frac{g_m^2}{1 + g_m^2 a_m^2}.
\end{equation}
The mutual information between the input and output of~\refeqn{eqn:channelFixedUncorrelated}, on the other hand, is quantified as
\begin{equation}\label{eqn:mutualInfoGaussian}
    \Isu{z_m;x_m|a_m} = \ln \brc{1 + g_m^2 a_m^2},
\end{equation}
so that using \eqref{eqn:mutualInfoParallel} we obtain the total achievable sum-rate \eqref{eqn:mutualInfoGeneric} of a single-user MIMO system with fixed diagonal channel.
\end{example}

\begin{example}[\emph{QPSK inputs}]
When the prior distribution of the elements of $\bs$ is given by $p(s) = 1/4\; \delta(s\pm 1/\sqrt{2} \pm \j/\sqrt{2})$, we have
\begin{align}\label{eqn:estimateMmseQpsk}
    \mmse{x_m} =&\; \frac{g_m}{\sqrt{2}} \tanh\brc{\sqrt{2} g_m a_m \;\re\{z_m\} } \nonumber\\
    &+ \j\frac{g_m}{\sqrt{2}} \tanh\brc{ \sqrt{2} g_m a_m\; \im\{z_m\} }.
\end{align}
Furthermore, the $(m,m)$th element of the MMSE matrix is given by
\begin{align}\label{eqn:matrixMmseGaussian}
    &\E_{z_m,x_m} \figbrc{|x_m-\mmse{x_m}|^2|a_m} \nonumber\\
    &= g_m^2-\frac{g_m^2}{\sqrt{2\pi}}\int_{\mathbb{R}}
    \tanh\brc{g_m^2 a_m^2 - g_m a_m\;s}\e^{-\frac{s^2}{2}}\d s,
\end{align}
and the per-stream mutual information is evaluated as
\begin{align}\label{eqn:mutualInfoQpsk}
    &\Isu{z_m; x_m|a_m} = 2 g_m^2 a_m^2  \nonumber\\
     &- \sqrt{\frac{2}{\pi}}\int_{\mathds{R}} \ln \cosh\brc{g_m^2 a_m^2 - g_m a_m\;s} \e^{-\frac{s^2}{2}} \d s.
\end{align}
\end{example}

\begin{example}[\emph{16-QAM inputs}]
When the elements of $\bs$, are uniformly drawn from the standard 16-QAM constellation, the diagonal terms of the MMSE matrix are evaluated as in~\refeqn{eqn:mmse16qam} on the top of the page (the minor typo in \cite[(27)~--~(28)]{lozano2006optimum} is corrected there).  The mutual information is then obtained numerically through the I-MMSE relation \cite{guo2005mutual}.
\end{example} 

\section{Asymptotic Achievable Sum-Rate}\label{sec:sumRate}
In this section, we present the main findings of the paper, namely, the asymptotic sum-rate of reliable communication over a multi-access MIMO channel in the presence of interferers. The expression is derived in the large-system limit (LSL), where the numbers of antennas at each terminal grow without bounds at constant ratios, \ie, $\betask{k} \Msk{k} = N \to \infty, \; \forall k \in\{1, \ldots, K\}$ and $\betail{\l} \Mil{\l} = N \to \infty, \; \forall \l \in\{1, \ldots, L\}$, where $\betask{k}$ and $\betail{\l}$ are finite positive constants.

In the remainder of the section, we first present the asymptotic result for a general (correlated) MIMO-MAC, and then specialize to the uncorrelated case where the expression are much simpler.

\subsection{General Case}
The main result of the paper is given in the following proposition\footnote{Even though some of the steps in the replica method are still lacking rigorous proof, the key results of the present paper are presented as propositions, being a convention in the replica calculus literature.}.

\begin{figure*}
\normalsize
\setcounter{equation}{31}
\begin{align}\label{eqn:mainResultUncorrelated}
    &\frac{1}{N}I(\by; \bxs)\nonumber = \sum_{k=1}^K \frac{1}{\betask{k}} \Isu{z_{\s,k}; x_{\s,k} \big| \sqrt{\rhosk{k}\xisk{k}}} + \sum_{\l=1}^L \frac{1}{\betail{\l}} \Isu{z_{\i,l}; x_{\i,\l} \big| \sqrt{\rhoil{\l}\xiil{\l}}} + \ln \brc{1 + \sum_{k=1}^K \epssk{k} + \sum_{\l=1}^L \epsil{\l}}\nonumber\\
    &\;- \sum_{k=1}^K \frac{1}{\betask{k}} \xisk{k} \epssk{k} - \sum_{\l=1}^L \frac{1}{\betail{\l}}\xiil{\l} \epsil{\l}  - \sum_{\l=1}^{L} \frac{1}{\betail{\l}} \Isu{\bar{z}_{\i,\l}; x_{\i,\l} \bigg| \sqrt{\rhoil{\l}\xiilbar{\l}}} - \ln \brc{1 +  \sum_{\l=1}^L\epsilbar{\l} } + \sum_{\l=1}^L \frac{1}{\betail{\l}} \xiilbar{\l} \epsilbar{\l}  + \calO\brc{\frac{1}{N}},
\end{align}
\hrulefill
\end{figure*}

\begin{proposition}\label{thm:miAsymptoticCorr}
Let the input distributions $p(\bxsk{k})$ and $p(\bxil{\l})$, as well as the spatial correlation matrices $\bTsk{k}$, $\bRsk{k}$, $\bTil{\l}$ and $\bRil{\l}$,
be given.  Then, the ergodic mutual information~\eqref{eqn:mutualInfo} normalized by the number of antennas at the receiver reads in the LSL as
\setcounter{equation}{27}
\begin{equation}\label{eqn:miAsymptotic}
    \frac{1}{N}I(\by; \bxs) = \hs - \hi + \calO\brc{\frac{1}{N}},
\end{equation}
where
\begin{align}
    \hs =&\; \frac{1}{N} \sum_{k=1}^K \Isu{\bzsk{k}; \bxsk{k} \big| \bAsk{k}}- \sum_{k=1}^K \frac{1}{\betask{k}} \xisk{k} \epssk{k}  \nonumber \\
    &\;+ \frac{1}{N}\sum_{\l=1}^L \Isu{\bzil{\l}; \bxil{\l} \big| \bAil{\l}} - \sum_{\l=1}^L \frac{1}{\betail{\l}}\xiil{\l} \epsil{\l}\nonumber \\
    &\;+ \frac{1}{N}\ln \det \! \brc{\!\bI_N\! +\! \sum_{k=1}^K \epssk{k}\bRsk{k}\! +\! \sum_{\l=1}^L \epsil{\l} \bRil{\l}\!}\! \nonumber\\
    &\;+1 + \ln \pi,\label{eqn:entropyYAsymptotic}\\
    \hi =&\; \frac{1}{N}\sum_{\l=1}^{L} \Isu{\barzil{\l}; \bxil{\l} | \barAil{\l}} - \sum_{\l=1}^L \frac{1}{\betail{\l}} \xiilbar{\l} \epsilbar{\l} \nonumber\\
    &\;+ \frac{1}{N}\ln \det \brc{\bI_N + \sum_{\l=1}^{L}\epsilbar{\l} \bRil{\l}} + 1 + \ln \pi. \label{eqn:entropyYConditionalAsymptotic}
\end{align}
The parameters $\epssk{k}$, $\xisk{k}$, $\epsil{\l}$, $\xiil{\l}$, $\epsilbar{\l}$ and $\xiilbar{\l}$ satisfy the following set of fixed-point equations\footnote{In general, the fixed-point equations may have more than one set of solutions. Among those, the one minimizing both~\refeqn{eqn:entropyYAsymptotic} and~\eqref{eqn:entropyYConditionalAsymptotic}, corresponding to entropy terms $h(\by|\calH)$ and $h(\by|\bxs,\calH)$, respectively. In physics, this phenomenon is referred to as \emph{phase coexistence}~\cite{nishimori2001statistical}. Note, however, that according to~\cite{nishimori2001comment}, one should expect the number of coexisting solutions to be finite since the phase space of a related problem in CDMA is known to be simple.}
\begin{subequations}\label{eqn:equationsFp}
\begin{align}
    \xisk{k} =&\; \frac{1}{\Msk{k}} \tr \figbrc{\bRsk{k} \sqbrc{\bI_{N} + \sum_{k=1}^{K} \epssk{k} \bRsk{k} + \sum_{\l=1}^{L} \epsil{\l} \bRil{\l}}^{-1} },\\
    \xiil{\l} =&\; \frac{1}{\Mil{\l}} \tr \figbrc{\bRil{\l} \sqbrc{\bI_{N} + \sum_{k=1}^{K} \epssk{k} \bRsk{k} + \sum_{\l=1}^{L} \epsil{\l} \bRil{\l}}^{-1} },\\
    \xiilbar{\l} =&\; \frac{1}{\Mil{\l}} \tr \figbrc{\bRil{\l} \sqbrc{\bI_{N} + \sum_{\l=1}^{L} \epsilbar{\l} \bRil{\l}}^{-1} },\label{eqn:xiBar}\\
    \epssk{k} =&\; \frac{\rhosk{k}}{\Msk{k}}  \tr\figbrc{ \mmsesu{\bxsk{k}, \bAsk{k}}\bTsk{k}}, \label{eqn:mmseUser}\\
    \epsil{\l} =&\; \frac{\rhoil{\l}}{\Mil{\l}}  \tr\figbrc{ \mmsesu{\bxil{\l}, \bAil{\l}}\bTil{\l} },\label{eqn:mmseInterf1}\\
    \epsilbar{\l} =&\; \frac{\rhoil{\l}}{\Mil{\l}}   \tr\figbrc{\mmsesu{\bxil{\l}, \barAil{\l}}\bTil{\l}},\label{eqn:epsBar}
\end{align}
\end{subequations}
where $\bAsk{k} = \sqrt{\rhosk{k}\xisk{k}}\sqTsk{k}$, $\bAil{\l} = \sqrt{\rhoil{\l}\xiil{\l}}\sqTil{\l}$, $\barAil{\l} = \sqrt{\rhoil{\l}\xiilbar{\l}}\sqTil{\l}$ and the MMSE matrices are obtained \via  \eqref{eqn:matrixMmse}.
\end{proposition}

\begin{IEEEproof}
In the mutual information expression of~\eqref{eqn:mutualInfo}, term~\eqref{eqn:entropyY} represents the sum-rate of an uplink system with all interferers considered as being desired users and Gaussian noise being the only source of disturbance. The asymptotic expression~\eqref{eqn:entropyYAsymptotic}, corresponding to this term can therefore be derived following the lines of~\cite{wen2007asymptotic}, where the sum-rate of a MIMO-MAC without interferers is considered. Hence, we omit the part of the proof related to~\eqref{eqn:entropyYAsymptotic} and present only the part related to~\eqref{eqn:entropyYConditionalAsymptotic} in the Appendix.

\end{IEEEproof}

Here the mutual information terms in~\refeqn{eqn:entropyYAsymptotic} and~\eqref{eqn:entropyYConditionalAsymptotic}, as well as terms~\eqref{eqn:mmseUser},~\eqref{eqn:mmseInterf1} and~\eqref{eqn:mmseInterf1}, are associated with two different fixed channels given by~\refeqn{eqn:channelFixed} with corresponding channel matrices $\bAsk{k}$, $\bAil{\l}$ and $\barAil{\l}$.
Hence, the terms $\epssk{k}$, $\epsil{\l}$ and $\epsilbar{\l}$ include the MMSE matrix of a fixed single-user channel discussed in Subsection~\ref{sec:channelFixedGeneral}, and transmit correlation of the original multiuser-MIMO channel \eqref{eqn:kronecker}. Despite looking a bit cumbersome, the two asymptotic expressions above have a simple interpretation. For instance,~\eqref{eqn:entropyYAsymptotic} represents the contributions of both users and interferers to the sum-rate of the MIMO-MAC presented in~\refeqn{eqn:signalReceived}. Meanwhile,~\eqref{eqn:entropyYConditionalAsymptotic} describes the amount of information discarded at the receiver due to noise and interference removal.

Here we emphasize the difference from the MAC system studied in~\cite{wen2007asymptotic}, where interferers were absent and white noise was the only source of disturbance. In contrast, our result, given in Proposition~\ref{thm:miAsymptoticCorr}, describes the sum-rate of the MIMO-MAC in the presence of interference. In the case of a single user and a single interferer, both using Gaussian signals, the above result immediately reduces to the mean mutual information reported in~\cite{moustakas2003mimo}, where it was obtained in a slightly different way. Note, though, that in contrast to~\cite{moustakas2003mimo}, our result is not restricted to Gaussian channel inputs, while the authors there computed also higher moments of mutual information.

\subsection{Uncorrelated Channels}
The next result provides the sum-rate for the special case where $\bTsk{k}$, $\bTil{\l}$, $\bRsk{k}$ and $\bRil{\l}$ are all identities.

\begin{corollary}\label{thm:mainResultUncorrelated}
In the LSL, the asymptotic average sum-rate of an uncorrelated MIMO-MAC given in~\emph{\refeqn{eqn:matrixChannel}} in the presence of interference is given by~\refeqn{eqn:mainResultUncorrelated} on the top of the page, where parameters $\epssk{k}$, $\xisk{k}$, $\epsil{\l}$, $\xiil{\l}$, $\epsilbar{\l}$ and $\xiilbar{\l}$ satisfy the following set of fixed-point equations
\setcounter{equation}{32}
\begin{subequations}\label{eqn:equationsFpUncorrelated}
\begin{align}
    \xisk{k} =&\; \betask{k}\brc{1 + \sum_{k=1}^{K} \epssk{k} + \sum_{\l=1}^{L} \epsil{\l}}^{-1},\\
    \xiil{\l} =&\; \betail{\l} \brc{1 + \sum_{k=1}^{K}  \epssk{k} + \sum_{\l=1}^{L}  \epsil{\l}}^{-1},\\
		\epssk{k} =&\; \frac{\rhosk{k}}{\Msk{k}}  \tr\figbrc{ \mmsesu{\bxsk{k}, \sqrt{\rhosk{k}\xisk{k}}\bI_{\Msk{k}}}},
	\label{eqn:mmseUserUncorrelated}\\
		\epsil{\l} =&\; \frac{\rhoil{\l}}{\Mil{\l}}  \tr\figbrc{ \mmsesu{\bxil{\l}, \sqrt{\rhoil{\l}\xiil{\l}}\bI_{\Mil{\l}}}},
		\label{eqn:mmseInterf1Uncorrelated}\\
    \xiilbar{\l} =&\; \betail{\l} \brc{1 + \sum_{\l=1}^{L} \epsilbar{\l}}^{-1},\\
		\epsilbar{\l} =&\; \frac{\rhoil{\l}}{\Mil{\l}}  \tr\figbrc{ \mmsesu{\bxil{\l}, \sqrt{\rhoil{\l}\xiilbar{\l}}\bI_{\Mil{\l}}}},\label{eqn:mmseInterf2Uncorrelated}
\end{align}
\end{subequations}
and the mutual information terms are obtained using \eqref{eqn:mutualInfoParallel}.
\end{corollary}

\begin{IEEEproof}
The proof follows directly from Proposition~\ref{thm:miAsymptoticCorr}. The result was also reported in our previous work~\cite{girnyk2012asymptotic}.
\end{IEEEproof}

\section{Precoder Optimization}\label{sec:optimizationCovariance}
As each transmitter has statistical CSI, by carefully choosing the precoder matrix, the transmitters could, in principle, maximize the sum mutual information between the inputs and outputs of their channels. The corresponding optimization problem is described as
\begin{equation} \label{opt:optimizationAsymptotic}
\begin{aligned}
    &\underset{\bcalGs}{\max}
    & & I(\by; \bxs)\\
    &\st
    & & \tr \{\bGsk{k}\bGskH{k}\} \leq \Msk{k},
    & & & k \in \calK \\
    & & & \bGsk{k}\bGskH{k} \succeq \bzero_{\Msk{k}},
    & & & k \in \calK,
\end{aligned}
\end{equation}
where $\calK \triangleq \{1,\ldots,K\}$, $\bcalGs \triangleq \{\bGsk{k}, \; \forall k \in \calK\}$  and the objective function is given in~\refeqn{eqn:mutualInfo}.  However, as mentioned in previous sections, working directly with the ergodic mutual information~\refeqn{eqn:mutualInfo} is difficult.  Thus, we next use the asymptotic results obtained in the previous section to simplify the optimization problem.

When examining Proposition~\ref{thm:miAsymptoticCorr}, we see that the random parts of the channels, $\bWsk{k}$ and $\bWil{\l}$, play no role in the mutual information when the system is sufficiently large.  Therefore, instead of the objective function~\refeqn{eqn:mutualInfo}, we maximize its asymptotic counterpart. Problem \eqref{opt:optimizationAsymptotic} then decouples into a set of individual per-transmitter optimization problems
\begin{equation} \label{opt:optimizationIndividual}
\begin{aligned}
    &\underset{\bGsk{k}}{\max}
    & & \Isu{\bzsk{k}; \bxsk{k} \big| \bAsk{k}}\\
    &\st
    & & \tr \{\bGsk{k}\bGskH{k}\} \leq \Msk{k},\\
    & & & \bGsk{k}\bGskH{k} \succeq \bzero_{\Msk{k}},
\end{aligned}
\end{equation}
where $\bAsk{k} = \sqrt{\rhosk{k}\xisk{k}}\sqTsk{k}$. Namely, each transmitter $k$ adjusts its own precoder matrix $\bGsk{k}$ according to its own transmit correlation matrix $\bTsk{k}$, which is available by the statistical-CSI assumption.

Note that here we need to maximize only the term $\hs$ in~\eqref{eqn:entropyYAsymptotic} since $\hi$ does not depend on the precoding matrices $\bGsk{k},\; \forall k\in\calK$.  On the other hand, the parameters $\epssk{k}$, $\xisk{k},\; \forall k$, and $\epsil{\l}$, $\xiil{\l}\; \forall \l$ do depend on the precoders in $\bcalGs$. To obtain a feasible point satisfying the KKT conditions~\cite{boyd2004convex}, we have to set to zero the derivatives of the objective w.r.t. the precoder matrices $\bGsk{k}$, given by
\begin{align}
    &\frac{1}{N}\nabla_{\bGsk{k}} I(\by;\bxs) = \nabla_{\bGsk{k}} \hs (\bGsk{k}) \nonumber\\
    & + \sum_{j=1}^{K} \frac{\partial \hs}{\partial \epssk{j}}  \nabla_{\bGsk{k}} \epssk{j}(\bGsk{k}) + \sum_{j=1}^{K} \frac{\partial \hs}{\partial \xisk{j}}  \nabla_{\bGsk{k}} \xisk{j}(\bGsk{k})\nonumber\\
    & + \sum_{\l=1}^{L} \frac{\partial \hs}{\partial \epsil{\l}} \nabla_{\bGsk{k}} \epsil{\l}(\bGsk{k}) + \sum_{\l=1}^{L} \frac{\partial \hs}{\partial \epsil{\l}} \nabla_{\bGsk{k}} \xiil{\l}(\bGsk{k}).
\end{align}
However, since $\hs$ represents the free energy given by an equation similar to~\eqref{eqn:freeEnergyEvaluated} in the Appendix, parameters $\epssk{k}$, $\xisk{k}$, $\epsil{\l}$ and $\xiil{\l}$ constitute its \emph{saddle-point}.  Therefore, the corresponding derivatives reduce to
(see also~\cite{wen2011asymptotic} for further discussion)
\begin{align}
    \frac{\partial \hs}{\partial \epssk{k}} = \frac{\partial \hs}{\partial \xisk{k}} = \frac{\partial \hs}{\partial \epsil{\l}} = \frac{\partial \hs}{\partial \epsil{\l}} = 0, \quad \forall k, \l
\end{align}
and it follows that
\begin{align}
    \nabla_{\bGsk{k}} I(\by;\bxs) = \nabla_{\bGsk{k}} \Isu{\bzsk{k}; \bxsk{k} \big| \bAsk{k}}, \quad \forall k.
\end{align}
Thus, for the sum-rate maximization we may consider $\epssk{k}$, $\xisk{k}$, $\epsil{\l}$ and $\xiil{\l}$ as being independent of $\bcalGs$.

Albeit optimization problem~\eqref{opt:optimizationIndividual} is, in general, non-convex, it can still be efficiently solved for the following (most practically relevant) special cases.

\subsection{Gaussian Inputs}
\label{sec:inputsGaussian}
In the case of Gaussian channel inputs, it is convenient to work with covariance matrices instead of precoders since the objective function of the optimization problem~\eqref{opt:optimizationIndividual} reduces to
\begin{equation}
    \Isu{\bzsk{k}; \bxsk{k} \big| \bAsk{k}} = \ln\det\brc{\bI_M + \bAsk{k}\bPsk{k}\bAskH{k}}.
\end{equation}
Let the \emph{singular-value decomposition} (SVD) of the effective fixed channel be given by $\bAsk{k} = \bU_{\bAsk{k}} \bSigma_{\bAsk{k}} \bV_{\bAsk{k}}^{\H}$, where $\bU_{\bAsk{k}}$ and $\bV_{\bAsk{k}}$ are orthonormal matrices and $\bSigma_{\bAsk{k}} = \diag([\sigma_{1}(\bAsk{k}),\ldots,\sigma_{\Msk{k}}(\bAsk{k})]^{\T})$ is the matrix with the singular values on the diagonal. Given the solution to the fixed-point equations ($\xisk{k}$ and $\epssk{k}$), the optimal input covariance matrix is then given by the \emph{water-filling} solution~\cite{cover2012elements}
\begin{equation}
\label{eqn:waterFilling}
    \bPsk{k}^{\star} = \bV_{\bAsk{k}} \bSigma_{\bPsk{k}} \bV_{\bAsk{k}}^{\H},
\end{equation}
where $\bSigma_{\bPsk{k}}$ is a diagonal matrix whose non-zero entries are
\begin{equation}
    [\bSigma_{\bPsk{k}}]_{m,m} = \sqbrc{\frac{1}{\nu} - \frac{1}{\sigma_{m}(\bAsk{k})}}^+,
\end{equation}
where $\nu$ is chosen so that the power constraint $\tr\{\bPsk{k}\} = \Msk{k}$ is satisfied.

We remark here that, as pointed out in~\cite{couillet2011deterministic}, to obtain the optimal transmit covariance matrix one has to iterate the solution to the fixed-point equations with the above statistical water-filling until the stopping criterion is reached.

\subsection{Discrete Inputs}
\label{sec:inputsDiscrete}
Unlike the previous case, finding the optimal precoder for discrete  constellations is a difficult task.  For such cases~\refeqn{opt:optimizationIndividual} is no longer a convex optimization problem.  It has been shown in~\cite{lamarca2009linear, xiao2011globally} that the mutual information is a concave function in the quadratic form $\bFsk{k} \triangleq \bAsk{k}\bGsk{k}\bGskH{k}\bAskH{k}$; yet, due to the power constraint, $\tr\{\bGsk{k}\bGskH{k}\} = \Msk{k}$, one cannot directly apply convex optimization tools for solving the problem. For instance, when using the \emph{gradient ascent} method for solving~\refeqn{opt:optimizationIndividual}, one updates $\bFsk{k}$ iteratively as
\begin{equation}
    \bFsk{k}^{(l+1)} = \bFsk{k}^{(l)} + \mu_{\bF} \Delta \bFsk{k},
\end{equation}
with $\mu_{\bF}$ being the step size and $\Delta \bFsk{k}$ being the gradient of $\Isu{\bzsk{k}; \bxsk{k} \big| \bAsk{k}}$ w.r.t. $\bFsk{k}$. It is shown in~\cite{palomar2006gradient} that the gradient of the mutual information in the single-user setup is
\begin{equation}
\label{eqn:miGradient}
    \nabla_{\bFsk{k}} \Isu{\bzsk{k}; \bxsk{k} \big| \bAsk{k}} = \bEsk{k},
\end{equation}
where we denoted for notational convenience $\bEsk{k}\triangleq\mmsesu{\bxsk{k},\bAsk{k}}$ for the MMSE matrix defined in~\refeqn{eqn:matrixMmse}. Nonetheless, in practice, precoder matrix $\bGsk{k}$ is updated subject to a power constraint, which limits its feasible region and complicates the problem.

Here we apply an algorithm similar to that proposed in~\cite{lamarca2009linear, xiao2011globally}, based on the alternating optimization between the following two subproblems.

\subsubsection{Per-Eigenmode Power Allocation}
Let the SVD of the precoder matrix be given by $\bGsk{k} = \bU_{\bGsk{k}} \bSigma_{\bGsk{k}} \bV_{\bGsk{k}}^{\H}$.
For fixed $\bV_{\bGsk{k}}$ the first subproblem is
\begin{equation} \label{opt:powerAllocation}
\begin{aligned}
    &\underset{\bSigma_{\bGsk{k}}^2}{\max}
    & & I\brc{\bzsk{k}; \bxsk{k} \big| \bAsk{k},\psu}\\
    &\st
    & & \tr \{\bSigma_{\bGsk{k}}^2\} \leq \Msk{k},\\
    & & & \bSigma_{\bGsk{k}}^2 \succeq \bzero_{\Msk{k}}.
\end{aligned}
\end{equation}
According to~\cite{payaro2009hessian}, this problem is convex, provided that the precoder has an optimal structure.

Since the matrix of interest, $\bSigma_{\bGsk{k}}^2$, is a diagonal matrix, we introduce for notational convenience a vector $\bg_k$, such that $\bSigma_{\bGsk{k}}^2 = \diag(\bg_k)$. We then choose an initial value for $\bg_k$, \eg, $\bg_k^{(1)} = 1/\Msk{k} \; \bones{\Msk{k}}$, and perform the gradient update\footnote{We remind the reader that $\Diag(\bA)$ denotes a column vector containing the diagonal entries of matrix $\bA$, whereas $\diag(\ba)$ denotes a diagonal matrix containing the entries of vector $\ba$.}
\begin{equation}
    \bg_k^{(l+1)} = \bg_k^{(l)} + \mu_{\bg} \brc{\Diag(\bSigma_{\bGsk{k}}^2 \bV_{\bGsk{k}}^{\H} \bEsk{k} \bV_{\bGsk{k}})-\gamma \bones{\Msk{k}}},
\end{equation}
with $\gamma = 1/\Msk{k}\bonesT{\Msk{k}}\Diag(\bSigma_{\bGsk{k}}^2 \bV_{\bGsk{k}}^{\H} \bEsk{k} \bV_{\bGsk{k}})$ and $\mu_{\bg}$ being an appropriately chosen step size, \eg, obtained by the \emph{backtracking line search} algorithm~\cite{boyd2004convex}. If $\bg_k^{(l+1)}$ has negative entries, one sets those to zero and renormalizes $\bg_k^{(l+1)}$ so that the power constraint is satisfied and then sets $\bSigma_{\bGsk{k}}^2 = \diag(\bg_k^{(l+1)})$.

\subsubsection{Optimization of the Eigenvectors of $\bFsk{k}$}
In this subproblem, for fixed $\bSigma_{\bGsk{k}}^2$ we optimize the eigenvectors of the quadratic form $\bFsk{k} = \bAsk{k}\bGsk{k}\bGskH{k}\bAskH{k}$. Let $\bLam_{\bF,k}$ be the diagonal matrix, whose entries are eigenvalues of $\bFsk{k}$. The second subproblem is then formulated as
\begin{equation} \label{opt:eigenvectorOptimization}
\begin{aligned}
    &\underset{\bFsk{k}}{\max}
    & & I\brc{\bzsk{k}; \bxsk{k} \big| \bAsk{k}, \psu}\\
    &\st
    & & \bLam_{\bFsk{k}} = \bSigma_{\bGsk{k}}^2 \bSigma_{\bAsk{k}}^2.
\end{aligned}
\end{equation}
The gradient of the mutual information is given by~\refeqn{eqn:miGradient}, and hence the gradient update for $\bFsk{k}$ becomes
\begin{equation}
    \bFsk{k}^{(l+1)} = \bFsk{k}^{(l)} + \mu_{\bF} \bEsk{k}.
\end{equation}
The obtained update has to be further projected into a matrix with the prescribed eigenvalues, which is as close to $\bFsk{k}^{(l+1)}$ as possible~\cite{lamarca2009linear,xiao2011globally}.

\section{Numerical Results}\label{sec:numericalResults}
\begin{figure}
\centering
\includegraphics[height = 7cm,width=8.5cm]{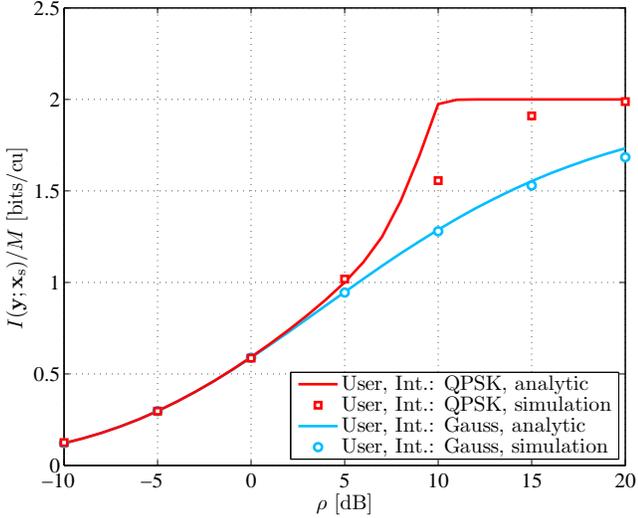}
  \caption{Average mutual information per dimension \vs SNR for the single-user single-interferer scenario. Both, the user and interferer, have the same type of signaling. The terminals are equipped with $M = N = 4$ antennas. Solid curves denote analytic results, markers denote the results of Monte-Carlo simulation.}
  \label{fig:resultMiInterfMonteCarlo}
\end{figure}

In this section, we provide numerical results alongside with some discussion. For the simulations, the spatial correlation at the transmitter side is assumed to be generated by a uniform linear antenna array with \emph{Gaussian power azimuth spectrum}~\cite{wen2011asymptotic}. Hence, correlation matrices ($\bTsk{k}$ and $\bTil{\l}$) consist of entries given by
\begin{equation}\label{eqn:correlation}
    \sqbrc{\bT}_{a,b} = \frac{1}{2 \pi \delta^2} \int_{-\pi}^{\pi} \e^{2 \pi \j d_{\lambda}(a-b)\sin(\phi) - \frac{(\phi-\theta)^2}{2\delta^2}} \d \phi,
\end{equation}
where $d_{\lambda}$ is the nearest neighbor antenna spacing (in wavelengths $\lambda$), $\theta$ is the mean angle and $\delta^2$ is the mean-square angle spread. For the sake of simplicity, we assume that there is no correlation at the receiver side, that is, $\bRsk{k} = \bI_{\Msk{k}}, \; \forall k$ and $\bRil{\l} = \bI_{\Mil{\l}}, \; \forall \l$.

\subsection{Uncorrelated Channels}

To begin with, we complement the obtained expression~\refeqn{eqn:mainResultUncorrelated} for the uncorrelated case with Monte-Carlo simulations~\cite{tranter2003principles}. We consider the setup, where a single user transmits its signal towards the receiver in the presence of a single interferer. All terminals have equal numbers of antennas, that is, $N = \Ms = \Mi = M$. Both the user and interferer utilize the same type of signals (either Gaussian or QPSK), and have the same total transmit power, that is, $\rhos = \rhoi = \rho$. In Fig.~\ref{fig:resultMiInterfMonteCarlo}, we plot the average mutual information per transmit antenna in bits per channel use (cu) as a function of SNR. Both the asymptotic results obtained \via the replica method and Monte Carlo simulations for $M = 4$ antennas are shown.
For QPSK, the simulations and asymptotic results are the farthest apart at SNRs around $\rho = 10$~dB due to the \emph{phase transition} phenomenon.
Namely, in this region the system instantly switches from one state to another, mimicking the ``water-ice'' transition in physics~\cite{muller2003channel}.  For Gaussian inputs, the plotted curve does not experience a phase transition and the asymptotic results are accurate already for small numbers of antennas.

\begin{figure}
\centering
\includegraphics[height = 7cm,width=8.5cm]{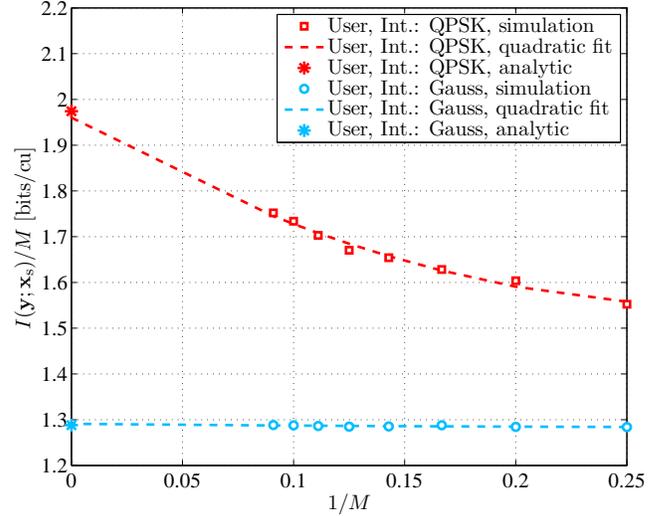}
  \caption{Average mutual information per dimension \vs the inverse of the number of antennas $M = N \in \{4,\ldots,11\}$ at terminals for both Gaussian and QPSK signaling schemes at SNR $\rho = 10$ dB. The asterisk markers at $1/M \to 0$ denote the predictions obtained by the replica analysis in the LSL.}
  \label{fig:resultMiInterfFit}
\end{figure}

To illustrate how the small scale simulations converge to the asymptotic result obtained using the replica method, Fig.~\ref{fig:resultMiInterfFit} plots the simulated values of the mutual information~\refeqn{eqn:mutualInfo} \vs $1/M$ for $M \in \{4,\ldots,11\}$ at $\rho = 10$ dB.
The markers at $1/M = 0$ represent the results obtained using Corollary~\ref{thm:mainResultUncorrelated} and quadratic curves are fitted to the simulated data using non-linear least-squares regression.  From the extrapolation we observe that the simulated per-antenna mutual information approaches close to the value predicted by the replica analysis also in the  region nearby the phase transition.

\begin{figure}
\centering
\includegraphics[height = 7cm,width=8.5cm]{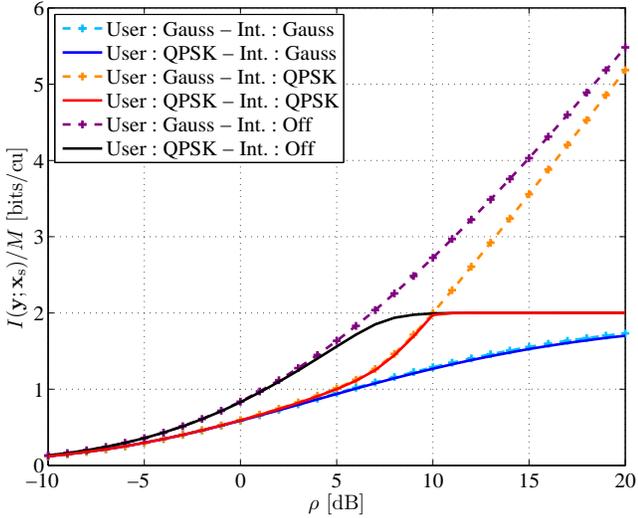}
  \caption{Average mutual information per dimension \vs SNR for different combinations of user's and interferer's signaling schemes. The terminals are equipped with $M = N = 4$ antennas.}
  \label{fig:resultMiInterfDifferent}
\end{figure}

Next, we consider the effect of signal constellations on the achievable sum-rate.  Fig.~\ref{fig:resultMiInterfDifferent} depicts the mutual information of the desired user when the Gaussian and/or QPSK signaling are used by the terminals. The interference-free case is also drawn for comparison. We directly see that for the desired user it is always best to employ Gaussian signaling. On the other hand, Gaussian signaling, when used by the interferer, creates more disturbance. Hence, in a cellular system where inter-cell interference is present, the network might be able to handle more users if some of them are assigned discrete constellations.  This is due to the fact that the most severe (unoptimized) interference is in fact Gaussian\footnote{Note that here we do not consider the optimization of the interferer's signal constellation aiming to jam the user. In the latter case, Gaussian signaling would not cause the worst-case interference, whereas an optimized discrete signaling would degrade the user's performance the most~\cite{shamai1992worst}.} \cite{cahn1971worst}.

\begin{figure}
\vspace{0.5cm}
\centering
\includegraphics[height = 7cm,width=8.5cm]{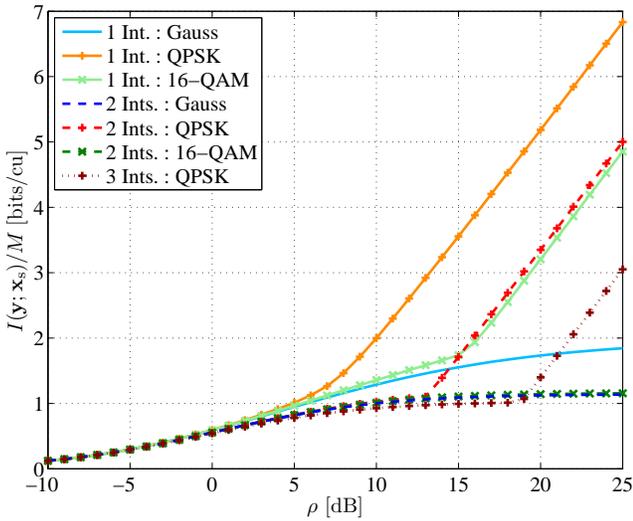}
  \caption{ Average mutual information per dimension \vs SNR for a single user with Gaussian signaling and $L \in\{1,2,3\}$ interferers using different signaling schemes. The terminals have $M = N = 4$ antennas.}
  \label{fig:resultMiInterfNumber}
\end{figure}

Fig.~\ref{fig:resultMiInterfNumber} presents the average mutual information per transmit antenna of a single user in the presence of $L \in \{1,2,3\}$ interferers using different signaling schemes (Gaussian, QPSK and 16-QAM). We see that $L = 2$ interferers using QPSK constellations create roughly the same performance degradation as a single interferer with 16-QAM signaling at high SNR.  On ther other hand, at SNR higher than 20 dB, $L = 3$ interferers with QPSK create smaller performance degradation than $L = 2$ interferers with 16-QAM. Again, we see that Gaussian signaling causes the worst-case degradation in the desired user's performance.

\subsection{Correlated Channels}

\begin{figure}
\centering
\includegraphics[height = 7cm,width=8.5cm]{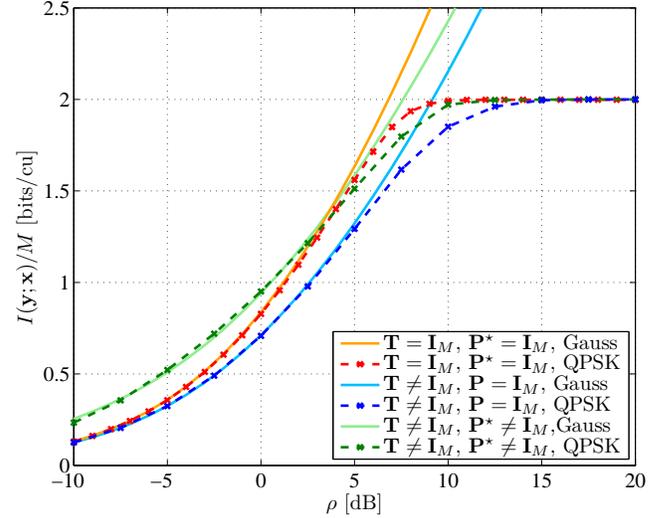}
  \caption{Average mutual information per dimension \vs SNR for a single-user system without interference.  Both correlated ($\bT \neq \bI_M$) and i.i.d.\ ($\bT = \bI_M$) MIMO channels with ($\bP \neq \bI_M$) and without ($\bP = \bI_M$) precoding are considered.  Gaussian or QPSK signaling is employed by the terminals that each are equipped with $M = N = 3$ antennas.}
  \label{fig:resultPrecoding}
\end{figure}

\begin{figure}
\centering
\includegraphics[height=7cm,width=8.5cm]{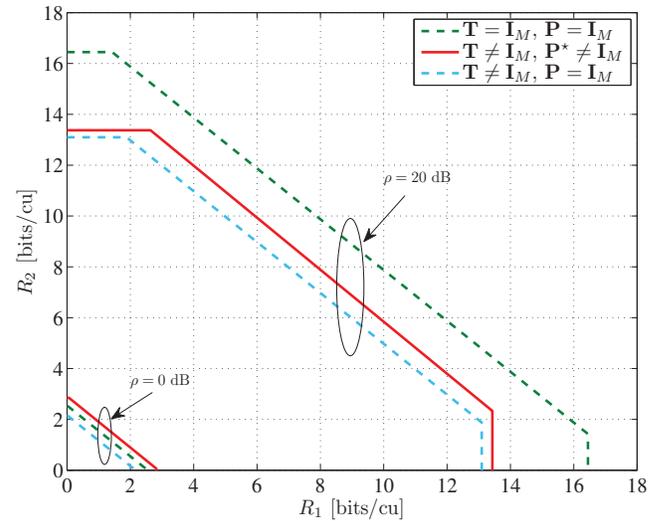}
  \caption{Achievable rate region for the 2-user MIMO-MAC under a power constraint of $\rho \in \{0,20\}$ dB.   Both correlated ($\bT \neq \bI_M$) and i.i.d. ($\bT = \bI_M$) channels with ($\bP \neq \bI_M$) and without ($\bP = \bI_M$) precoding are considered. The terminals have $M = N = 3$ antennas.}
  \label{fig:resultRateRegionMac}
\end{figure}

\begin{figure*}
\centering
\subfigure[Gaussian channel inputs.]{
\includegraphics[height = 7cm,width=8.5cm]{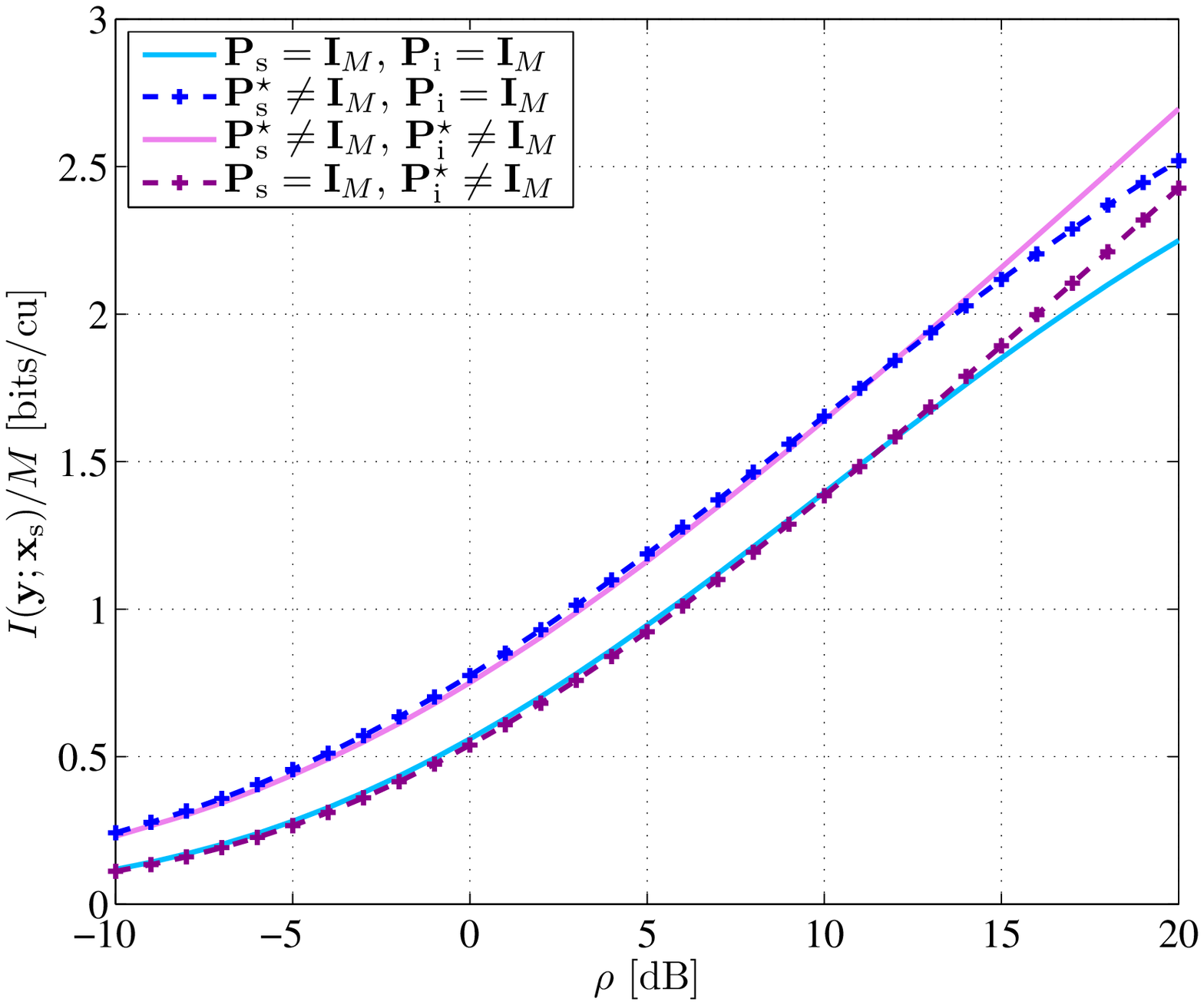}
\label{fig:resultPrecodingInterfGauss}
}
\subfigure[QPSK channel inputs.]{
\includegraphics[height = 7cm,width=8.5cm]{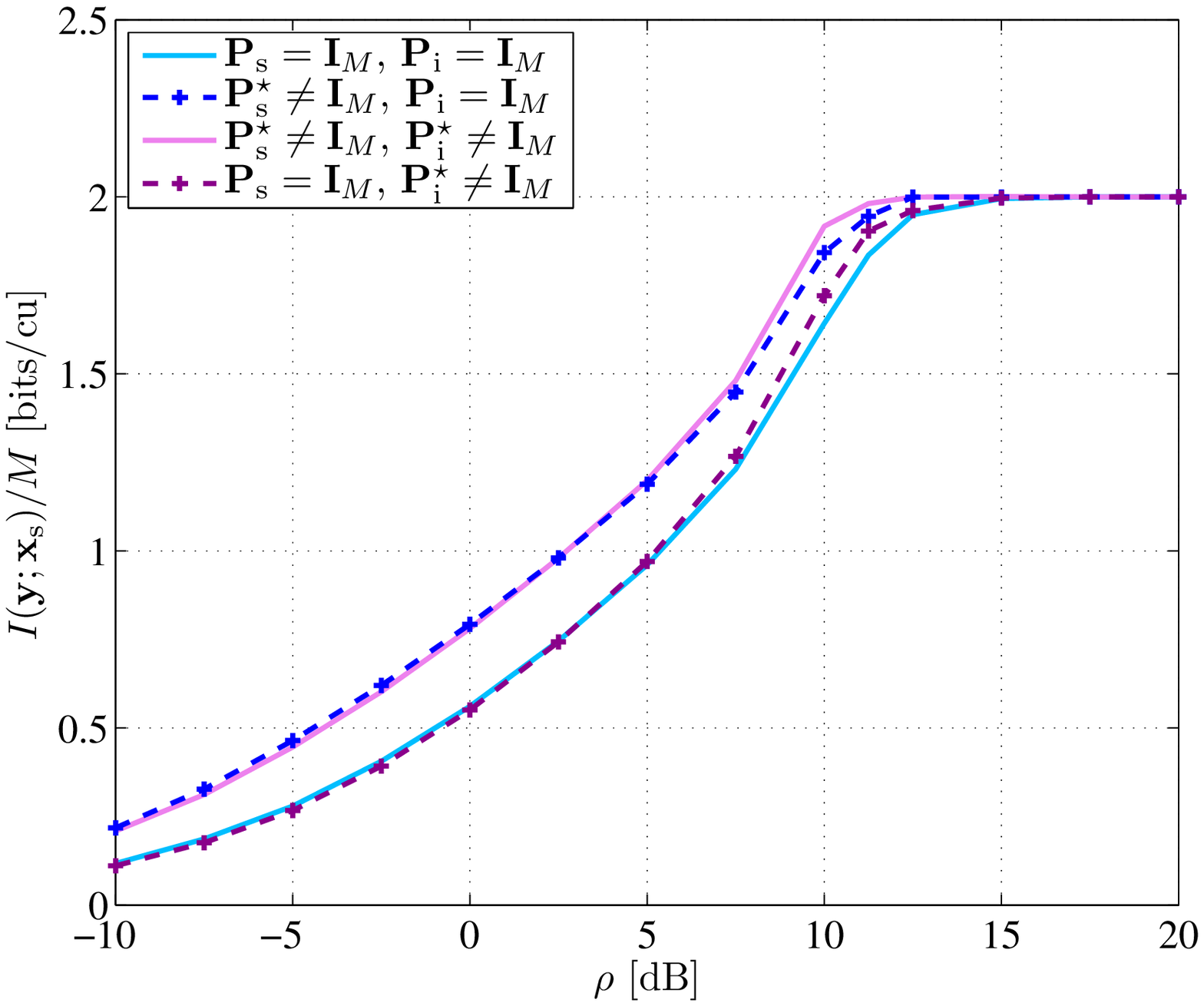}
  \label{fig:resultPrecodingInterfQpsk}
}
\caption[Effect of precoding for different signaling schemes]{Average mutual information per dimension \vs SNR for a single-user correlated MIMO channel with ($\bPs \neq \bI_M$) and without ($\bPs = \bI_M$) precoding.   A single interferer using precoded ($\bPi \neq \bI_M$) or isotropic ($\bPi = \bI_M$) channel inputs is present. The terminals have $M = N = 3$ antennas.}
\label{fig:resultPrecodingInterf}
\end{figure*}
In this section, we study the behavior of the system under spatial correlation and quantify the gains of precoding. Fig.~\ref{fig:resultPrecoding} depicts the normalized ergodic mutual information given Gaussian and QPSK inputs as a function of SNR of a single-user (no interference) MIMO channel with $N = \Ms = M = 3$ antennas under various conditions. Namely, we consider the cases of correlated and uncorrelated channels with and without precoding at the transmitter. The transmit side correlation parameters are set to the terminals as follows. The antenna spacing is set to $d_{\lambda} = 1$, the mean angle is set to $\theta = 0^{\circ}$ and the root-mean-square angle spread is chosen to be $\delta = 5^{\circ}$. The receive side correlation matrix is set to identity, \ie, $\bR = \bI_N$. As expected, at low SNR the curves representing the two constellations coincide. Moreover, for the case of Gaussian inputs, transmit correlation decreases the achievable rate at high SNR regardless of precoding. Quite remarkably though, at low SNR transmit correlation together with precoding based on the statistical water-filling~\refeqn{eqn:waterFilling} is beneficial in terms of the mutual information. Somewhat similar behavior is observed for the case of QPSK signals. At low SNR a precoder in combination with transmitter-side correlation allows to improve the system performance as compared to the case of an uncorrelated channel. However, since in this case the per-stream mutual information saturates at 2 bits/cu at high SNR, transmit correlation does not affect the rates too much in this region. To optimize the precoder matrix for the case of QPSK signals, we have used the algorithm described in Subsection~\ref{sec:inputsDiscrete}.

Next, we investigate the performance of a correlated MIMO-MAC. The rate region of a generic $K$-user MIMO-MAC using Gaussian signaling is given by~\cite{goldsmith2003capacity}
\begin{align} \label{eqn:regionCapacity}
    \calC&_{\textrm{MAC}} = \bigcup_{\substack{\tr\{\bPsk{k}\}\leq\Msk{k}\\ \bPsk{k} \succeq \bzero_{\Msk{k}}\\ \forall k \in\calK}} \Bigg\{ \{R_k\}, \forall k \in \calK : \nonumber\\
     & \sum_{i\in \calS} R_i \leq \ln\det\brc{\bI_N + \sum_{i\in \calS}\bHsk{i} \bPsk{i} \bHskH{i}}, \forall \calS \subset \calK \Bigg\}.
\end{align}
Note that the corresponding large-system ergodic mutual information terms can be directly obtained from Proposition~\ref{thm:miAsymptoticCorr}. To illustrate this region, we consider a symmetric setup with two users who both have $M = \Msk{1} = \Msk{2} = 3$ antennas.  We further fix the available transmit powers $\rho = \rhosk{1} = \rhosk{2}$ to $\rho \in \{0, 20\}$ dB and evaluate the achievable rate regions for the given 2-user MAC. The result is depicted in Fig.~\ref{fig:resultRateRegionMac} where both uncorrelated and correlated channels with and without precoding are present. It is clear that using precoding at both terminals is beneficial when transmit correlation is present. As expected though, at high SNR the rate region is largest for the uncorrelated MAC. On the contrary, at low SNR the largest rate region is achieved in the presence of correlation and optimal precoding.

To finish this section, we return to the case of one desired user and add an interferer with $\Mi = 3$ antennas, having the same transmit power, $\rhoi = \rho$, and same correlation parameters. Fig.~\ref{fig:resultPrecodingInterf} depicts the average mutual information as a function of SNR for this scenario under Gaussian and QPSK signaling schemes. Both the user and interferer either do or do not realize precoding. Note that the scenario is symmetric and hence the precoders used by the terminals are the same.
Moreover, the terminals adapt to their own correlation matrices aiming to increase their own rates. From Fig.~\ref{fig:resultPrecodingInterfGauss} we see that, quite expectedly, for the case of Gaussian signals, precoding at the user increases its own ergodic rate. At the same time, we also see that utilizing the optimal precoder at the interferer results in higher rate at the user's terminal at high SNR, degrading the performance of the latter in the low-SNR region only slightly. This observation falls along the lines of~\cite{lozano2003multiple}, where spatially colored noise was shown to be less harmful than white Gaussian noise. Interestingly, similar behavior is observed for the case of QPSK inputs (\cf Fig.~\ref{fig:resultPrecodingInterfQpsk}), apart from expected saturation at 2 bits/cu at high SNR.

\section{Conclusions}\label{sec:conclusion}
In this paper, we derived an explicit expression for the asymptotic achievable sum-rate of the MIMO multiple-access channel in the presence of interference. The result accounts for the spatial correlation at the terminals and, in contrast to the previous results, is not restricted to Gaussian signals. Although derived in the large system limit, it approximates relatively well the achievable sum-rate of small systems. We have also studied the impact of the number of interferers, their signaling scheme and spatial correlation structure on the system performance. For instance, Gaussian signaling is seen to create the worst-case (unoptimized) interference. Thus, the system may handle more interferers if they use discrete signal constellation, as compared to the case of Gaussian interferers. The obtained large-system approximation has been further used to find precoder matrices for maximizing the sum-rate for both Gaussian and finite-alphabet signaling schemes. It has been demonstrated that properly optimized precoder significantly increases the achievable rates. More interestingly, in the low-SNR region the presence of spatial correlation, in combination with an optimal precoder, is beneficial and can in fact improve the system performance as compared to the case of uncorrelated channels. The proposed approach is general and degenerates to many well-known results as special cases.

\appendix
\section{Proof of Proposition~\ref{thm:mainResult}}
In general, direct computation of~\refeqn{eqn:entropyYConditional} is very difficult if we allow arbitrary channel input distributions. To overcome this obstacle we use the replica method to compute the entropy in the LSL.  To stay coherent with the existing work, we partially keep the statistical physics terminology, avoiding unnecessary jargon whenever possible.

Let us define the \emph{partition function} related to~\refeqn{eqn:entropyYConditional} as
\begin{equation}\label{eqn:partitionFunctionDef}
    Z(\by,\bxs,\calH) \triangleq \E_{\bxi}\figbrc{\frac{1}{\pi^N}\e^{-\|\by-\bHs \bxs - \bHi \bxi\|^2}}.
\end{equation}
In statistical physics, virtually all interesting macroscopic quantities can be derived from the partition function of the system.  Often, however, it is more convenient to work with the logarithm of the partition function, or \emph{free energy}, instead.  If we further average the (normalized) free energy w.r.t.\ the remaining randomness in the MIMO setup \eqref{eqn:partitionFunctionDef}, we get
\begin{equation}\label{eqn:freeEnergyDef1}
    \calF \triangleq -\frac{1}{N}\E_{\by,\bxs,\calH} \ln Z\brc{\by, \bxs, \calH},
\end{equation}
that is, just the normalized equivocation term, $\frac{1}{N}h(\by|\bxs,\calH)$.  Then, we take the first step towards making the evaluation of $\calF$ solvable by writing
\begin{equation}\label{eqn:identity2}
    \calF = -\frac{1}{N}\lim_{u \rightarrow 0^+} \frac{\partial}{\partial u} \ln \E_{\by,\bxs,\calH} \figbrc{Z^u\brc{\by,\bxs,\calH}},
\end{equation}
and implicitly assuming that the system size also grows large as discussed in Section~\ref{sec:sumRate}.
This identity is exact when $u$ is a real number, but on its own it does not solve the problem.  Thus, we invoke the \emph{replica trick} and
write the under-log term as
\begin{align}\label{eqn:replicate2}
    &\E_{\by,\bxs,\calH} \figbrc{Z^u\brc{\by,\bxs,\calH}} \nonumber\\
    &=\E_{\bxs,\bXi,\calH} \figbrc{\int \frac{1}{\pi^N} \prod_{a=0}^u \e^{-\|\by - \bHs \bxs - \bHi \bxia{a}\|^2}  \d\by},
\end{align}
where $\bxia{a}$ is the $a$th replica of the signal vector transmitted by the interferers. Its distribution, $p(\bxia{a})$, is identical to $p(\bxi)$ and conditionally independent for $a\in\{0,1,\ldots,u\}$ given $\by$ and $\calH$. For ease of exposition, we have also defined the vector  $\bXi \triangleq [\bxiaT{0}, \ldots, \bxiaT{u}]^{\T} \in \Complex{L\Mi (u+1)}$ that contains the replicated interferers' signals. Note that since the partition function \eqref{eqn:partitionFunctionDef} has an expectation over $\bxi$ and not $\bxs$, we introduce only the replicas related to the former.

After applying the replica trick, the problem of finding the free energy reduces to evaluating $\E_{\by, \bxs,\calH} \{Z^u(\by, \bxs,\calH)\}$ for integer-valued $u$ using techniques from large deviations theory and then assuming that the result generalizes to real positive values, at least in the vicinity of zero\footnote{Note that mathematical rigor of this step is still an open problem. However, some results obtained by the replica method are confirmed to match the ones derived \via systematic approaches (\eg,~\cite{hachem2008new},~\cite{nishimori2001statistical}). Moreover, the results can be further verified \via Monte-Carlo simulations, as we saw in Section~\ref{sec:numericalResults}. Therefore, we regard the replica analysis as a valid mathematical tool.}.

Let us now define the following set of random vectors
\begin{subequations}
\begin{align}
    \bvsk{k} & \triangleq  \sqrt{\frac{\rhosk{k}}{\Msk{k}}} \bHsk{k} \bxsk{k} \in \Complex{N},\\
    \bvila{\l}{a} & \triangleq \sqrt{\frac{\rhoil{\l}}{\Mil{\l}}} \bHil{\l} \bxila{\l}{a} \in \Complex{N}.
\end{align}
\end{subequations}
Denote also $\bvs  \triangleq \sum_{k=1}^K \bvsk{k}$ and $\bvia{a}  \triangleq \sum_{\l=1}^L \bvila{\l}{a}$, and group them into a concatenated vector
\begin{equation}\label{eqn:vectorV}
    \bV \triangleq [\bvsT + \bviaT{0}, \ldots, \bvsT + \bviaT{u}]^{\T} \in \Complex{N (u+1)}.
\end{equation}  
Conditioned on the interferers' signals $\bXi$, we know by the central limit theorem that as the dimensions of the channel matrices $\bHsk{k}$ and $\bHil{\l}$ grow large, $\bV$ converges to a zero-mean Gaussian random vector with conditional covariance
\begin{equation}\label{eqn:covariance}
    \bQ = \sum_{k=1}^K (\bQsk{k}\otimes\bRsk{k}) + \sum_{\l=1}^L (\bQil{\l}\otimes\bRil{\l}).
\end{equation}
The auxiliary matrix $\bQil{\l}$ has entries
\begin{align}
    [\bQil{\l}]_{a,b} &= \frac{\rhoil{\l}}{\Mi} \bxilaH{k}{b} \bTil{\l} \bxila{\l}{a},   \label{eqn:Qilab}
\end{align}
for $a,b \in \{0,1,\ldots,u\}$, while
$\bQsk{k} = \qsk{k} \bones{u+1} \bonesT{u+1}$ with
\begin{equation}
\qsk{k} = \frac{\rhosk{k}}{\Ms} \bxskH{k} \bTsk{k} \bxsk{k}.
\end{equation}
Note that here we used~\eqref{eqn:kronecker} and the assumption that $\bWsk{k}$ and $\bWil{\l}$ have i.i.d. CSCG entries of unit variance to derive the result.

\begin{figure*}[t!]
\normalsize
\setcounter{equation}{66}
\begin{subequations}\label{eqn:termsT}
\begin{align}\label{eqn:firstTermT}
    \funcT{1}(\calQ,\tcalQ) =&\;  u\ln\det\Bigg(\bI_{N(u+1)}
    +  \sum_{\l=1}^{L}(\pil{\l}-\qil{\l})\bRil{\l}\Bigg),\\
    \label{eqn:secondTermT}
    \funcT{2}(\calQ,\tcalQ) =&\; \sum_{k=1}^K \Msk{k}(u+1)(\tilpsk{k} + u \tilqsk{k})\qsk{k}
    + \sum_{\l=1}^L \Mil{\l} (u+1)(\tilpil{\l} \pil{\l} + u \tilqil{\l} \qil{\l}),\\
		\funcT{3}(\calQ,\tcalQ)
    =&\; \sum\limits_{k=1}^{K} \Msk{k} \ln \brc{1 - u(u+1)\barAsk{k}^2} -\sum\limits_{\l=1}^{L}\ln  \E_{\bXi} \Bigg\{ \exp \Bigg( \bigg\|\sum\limits_{a=0}^{u}\barAil{\l} \bxila{\l}{a}\bigg\|^{2}
    - \sum\limits_{a=0}^{u} \bxilaH{\l}{a} (\barAil{\l}^2 - \barBil{\l}) \bxila{\l}{a} \Bigg) \Bigg\} \label{eqn:termLast}
\end{align}
\end{subequations}
\hrulefill
\end{figure*}
\setcounter{equation}{58}

Let us define $\calQ \triangleq \{(\bQsk{k}, \bQil{\l}),\; \forall k,\l\}$, so that the expectation over replicated vectors $\bXi$ may be rewritten as an integral over a probability measure of $\calQ$.  Treating $\bV$ as a Gaussian random vector, it can be shown \via the Edgeworth expansion that in the
LSL~\cite{tanaka2002statistical} 
\begin{align}\label{eqn:integralOverMeasure}
    \E_{\by,\bxs,\calH} \{Z^u(\by,\bxs,\calH)\}
    =&  \int \e^{\funcG(\calQ)} d\measure(\calQ),
\end{align}
where we have omitted constant terms and $\measure(\calQ)$ reads
\begin{align}\label{eqn:measure}
    \measure(\calQ)
    & = \E \Bigg\{\prod_{k=1}^K \ind \brc{ \rhosk{k}\bxskH{k} \bTsk{k} \bxsk{k} - \Ms \qsk{k} } \nonumber\\
    & \times \prod_{a,b=0}^u \prod_{\l=1}^L \ind \brc{ \rhoil{\l} \bxilaH{k}{b} \bTil{\l} \bxila{\l}{a} -  \Mi [\bQil{\l}]_{a,b} }\Bigg\}.
\end{align}
with the above expectation being w.r.t. $\{\bXi,\calH\}$. If we plug $\bV$ into
\eqref{eqn:replicate2} and assess the expectations w.r.t.\ $\bV$ and $\by$ using Gaussian integration, we get
\begin{align}
\label{eqn:gFunction}
    &\funcG(\calQ)  = - Nu\ln\pi - N \ln (u+1) \nonumber\\
    & - \ln\det\sqbrc{\bI_{N(u+1)} + \sum_{k=1}^{K}\bQsk{k}\bSigma \otimes \bRsk{k} + \sum_{\l=1}^{L}\bQil{\l}\bSigma \otimes \bRil{\l} },
\end{align}
where $\bSigma \triangleq \bI_{u+1} - \frac{1}{u+1} \bones{u+1} \bones{u+1}^{\T} \in \Real{(u+1) \times (u+1)}$.

To compute the integral in~\eqref{eqn:integralOverMeasure}, we note that since both $\bQsk{k}$ and $\bQil{\l}$ are formed by summing independent random variables, measure~\eqref{eqn:measure} satisfies the large deviations property and by Varadhan's theorem~\cite{ellis2005entropy}
\begin{align}
\label{eqn:varadhanTheorem}
    \frac{1}{N}\ln \E_{\by,\bxs,\calH}& \{Z^u(\by,\bxs,\calH)\} \nonumber\\
    & - \frac{1}{N}\underset{\calQ}{\max}\brc{\funcG(\calQ) - \funcRate (\calQ)} \to 0,
\end{align} 
in the LSL. The second term inside the maximization is referred to as the rate function and can be obtained \via Cram\'{e}r's theorem~\cite{ellis2005entropy}
\begin{align}
    \label{eqn:rate}
    \funcRate(\calQ)  =&\; \underset{\tcalQ}{\max} \Bigg\{ \sum_{k=1}^K \Msk{k} \tr \{\tilQsk{k} \bQsk{k}\} \nonumber\\
     &+ \sum_{\l=1}^L \Mil{\l} \tr \{\tilQil{\l} \bQil{\l}\} - \ln \mgf(\tcalQ)\Bigg\},
\end{align}
where the moment-generating function of $\measure(\calQs,\calQi)$ reads
\begin{align}
    \label{eqn:mgf}
    \mgf(\tcalQ) =
    \E_{\bXi} \Bigg\{&\prod_{k=1}^K \e^{ \rhosk{k} \bXskH{k} (\tilQsk{k} \otimes \bTsk{k}) \bXsk{k}}  \nonumber\\
     & \times\prod_{\l=1}^L \e^{ \rhoil{\l} \bXilH{\l} \brc{\tilQil{\l} \otimes \bTil{\l}} \bXil{\l} } \Bigg\},
\end{align}
and we denoted $\bXsk{k}~\triangleq~[\bxskT{k}, \ldots, \bxskT{k}]^{\T} \in \Complex{\Msk{k}(u+1)}$, $\bXil{\l} \triangleq [\bxilaT{\l}{0}, \ldots, \bxilaT{\l}{u}]^{\T} \in \Complex{\Mil{\l}(u+1)}$.  As before, we group the auxiliary ``Q-matrices'' as $\tcalQ \triangleq \{(\tilQsk{k},\tilQil{\l}),\;\forall k,\l\}$.

To make the optimization problems in \eqref{eqn:varadhanTheorem}~and~\eqref{eqn:rate} tractable, we next assume that the saddle-point solutions are invariant under the permutation of the replica indices.  This is known as the \emph{replica symmetric} (RS) ansatz\footnote{This assumption has been widely accepted in the field of statistical physics~\cite{nishimori2001statistical} and information theory~\cite{muller2003channel},~\cite{guo2005cdma},~\cite{wen2007asymptotic}. However, the cases of \emph{replica-symmetry breaking} are known in the literature~\cite{zaidel2012vector},~\cite{muller2008vector}.} and here it implies that we can write the members of
$\calQ$ and $\tcalQ$ as
\begin{subequations}
\begin{align}
    \tilQsk{k} =& \tilqsk{k} \bones{u+1} \bonesT{u+1} + (\tilpsk{k} - \tilqsk{k}) \bI_{u+1},\\
    \bQil{\l} =& \qil{\l} \bones{u+1} \bonesT{u+1} + (\pil{\l} - \qil{\l}) \bI_{u+1},\\
    \tilQil{\l} =& \tilqil{\l} \bones{u+1} \bonesT{u+1} + (\tilpil{\l} - \tilqil{\l}) \bI_{u+1}.
\end{align}
\end{subequations}
Under the RS assumption, the free energy in the LSL becomes
\begin{equation}
\label{eqn:freeEnergyEvaluated}
    \mathcal{F} = 1+\ln \pi + \frac{1}{N}\lim_{u \rightarrow 0^+} \frac{\partial}{\partial u} \; \underset{\calQ}{\min} \; \underset{\tcalQ}{\max}\;
    \sum_{i=1}^{3} \funcT{i}(\calQ,\tcalQ),
\end{equation}
where the terms of $\funcT{i}(\calQ,\tcalQ)$ are given in~\eqref{eqn:termsT}, on the top of the page, where we denoted $\barAsk{k} \triangleq \sqrt{\rhosk{k} \tilqsk{k}} \bTsk{k}^{1/2}$, $\barAil{\l} \triangleq \sqrt{\rhoil{\l} \tilqil{\l}} \bTil{\l}^{1/2}$ and $\barBil{\l} \triangleq \rhoil{\l}\tilpil{\l} \bTil{\l}$.  We also assumed there that all terminals have independent channel inputs. Performing the Hubbard-Stratonovich transform~\cite{stratonovich1957method},~\cite{hubbard1959calculation} on~\refeqn{eqn:termLast}, we decouple the quadratic terms
\begin{align} \setcounter{equation}{67}
    \label{eqn:hubbardStratonovich}
    \funcT{3}& (\calQ,\tcalQ) \nonumber\\
    =&\; \sum\limits_{k=1}^{K} \Msk{k}\ln\det \brc{1 - u(u+1)\barAsk{k}^2} \nonumber\\
    & - \sum_{\l=1}^{L} \ln \frac{1}{\pi^{\Mi}} \int \E_{\bxil{\l}} \left\{ \e^{ -\|\barzil{\l} - \barAil{\l} \bxil{\l}\|^2} \e^{ \bxilH{\l} \barBil{\l} \bxil{\l}} \right\}\nonumber\\
    &\times\sqbrc{ \E_{\bxil{\l}} \figbrc{ \e^{ 2 \re\{\bzilH{\l} \barAil{\l} \bxil{\l}\} - \bxilH{\l} (\barAil{\l}^2 - \barBil{\l}) \bxil{\l}}  } }^u \d\bzil{\l},
\end{align}
where $\barzil{\l} \in \Complex{\Mil{\l}}$ is an auxiliary variable.

To solve~\refeqn{eqn:freeEnergyEvaluated}, we find the conditions under which  the derivatives of the argument w.r.t.\ all the RS parameters vanish. After taking $u \to 0$, we get 
\begin{subequations} \label{eqn:saddlePointConditions}
\begin{align}
    \tilpsk{k} =&\; \tilqsk{k} = \tilpil{\l} = 0, \quad \forall k,\l,\\
    \tilqil{\l} =&\; \frac{\rhoil{\l}}{\Mil{\l}} \tr \figbrc{\bS^{-1} \bRil{\l}},\\
    \pil{\l} -&\; \qil{\l} = \frac{1}{\Mil{\l}} \tr\figbrc{ \mmsesu{\bxil{\l}, \barAil{\l}}\bTil{\l}},
\end{align}
\end{subequations}
where we wrote for notational convenience
\begin{equation}
    \bS \triangleq \bI_{N} + \sum_{\l=1}^{L} (\pil{\l} - \qil{\l}) \bRil{\l},
\end{equation}
and the MMSE matrix is defined in \eqref{eqn:matrixMmse}.
Finally, taking the derivative w.r.t. $u$ and letting $u \to 0$ we get
\begin{align}
    \calF =&\; \ln \det \bS + \sum_{\l=1}^L \Isu{\barzil{\l}; \bxil{\l} \big| \barAil{\l}}\nonumber \\
    & + \sum_{\l=1}^L \Mil{\l}\tilqil{\l} (\qil{\l} -\pil{\l}) + 1+\ln \pi,
\end{align}
where we used the fact that $\pil{\l} = \rhoil{\l}$. Denoting $\xiilbar{\l} \triangleq \tilqil{\l}$ and $\epsilbar{\l} \triangleq \pil{\l} - \qil{\l}$, we obtain $\hi$ from~\eqref{eqn:entropyYConditionalAsymptotic}, as well as a system of fixed-point equations given by~\refeqn{eqn:epsBar} and~\eqref{eqn:xiBar}.

\section*{Acknowledgement}
The authors thank the anonymous reviewers for their suggestions that have greatly improved the quality of the manuscript. In addition, the authors are grateful to Peter Larsson for the discussions on practical considerations regarding the scenario investigated here.

\bibliographystyle{IEEEtran}
\bibliography{Bibliography/literature}

\end{document}